\newif\iftwocolumn
\newcommand{\figwidth}{\iftwocolumn \columnwidth \else 100mm\fi}

\documentclass[aps,prapplied,amsmath,amssymb,reprint,groupedaddress]{revtex4-2}\twocolumntrue

\usepackage[utf8]{inputenc}
\usepackage[english]{babel} 
\usepackage[T1]{fontenc}
\usepackage{graphicx}
\usepackage{placeins}
\usepackage{amsmath,amssymb}
\usepackage[colorlinks=true,citecolor=red]{hyperref}		
\usepackage{natbib}
\usepackage{xcolor}
\usepackage{siunitx}
\usepackage{braket}

\newcommand\ea{{\em et al.}}
\definecolor{groen}{RGB}{0,150,0}

\newcommand{\Rb}{$^{87}$Rb}

\begin{document}
\title{Reducing number fluctuations in an ultracold atomic sample using Faraday rotation and iterative feedback}
\author{R. Thomas$^1$}\email{ryan.j.thomas1@gmail.com}
\author{J. S. Otto$^2$}
\author{M. Chilcott$^2$} 
\author{A. B. Deb$^2$}
\author{N. Kj{\ae}rgaard$^2$}\email{niels.kjaergaard@otago.ac.nz}
\affiliation{$^1$Department of Quantum Science, Research School of Physics, The Australian National University, Canberra 2601, Australia}
\affiliation{$^2$Department of Physics, QSO-Centre for Quantum Science, and Dodd-Walls Centre, University of Otago, Dunedin, New Zealand}

\begin{abstract}
We demonstrate a method to reduce number fluctuations in an ultracold atomic sample using real-time feedback.  By measuring the Faraday rotation of an off-resonant probe laser beam with a pair of avalanche photodetectors in a polarimetric setup we produce a proxy for the number of atoms in the sample.  We iteratively remove a fraction of the excess atoms from the sample to converge on a target proxy value in a way that is insensitive to environmental perturbations and robust to errors in light polarization.  Using absorption imaging for out-of-loop verification, we demonstrate a reduction in the number fluctuations from $3\%$ to $0.45\%$ for samples at a temperature of $16.4$ \si{\micro\kelvin} over the time-scale of several hours which is limited by temperature fluctuations, beam pointing noise, and photon shot noise.

\end{abstract}

\maketitle
\section{Introduction}
\label{sec:Introduction}
The ability to isolate, trap, and cool samples of atoms to ultra-low temperatures ($\sim$$1$ \si{\micro\kelvin}) has allowed for the exploration of many questions in fundamental quantum science as well as the development of novel sensors with unprecedented precision and accuracy.  Ultracold atomic systems have been used to study soliton dynamics \cite{Burger1999,Everitt2017}, phase separation \cite{Papp2008,Ospelkaus2006}, super and sub-radiance \cite{Ferioli2020,Inouye1999}, magnetic order \cite{Sadler2006}, quantum phase transitions \cite{Luo2017,Baumann2010,Greiner2002}, and fundamental chemistry \cite{Ospelkaus2010,Croft2017,Hu2014}.  Ultracold atoms are also the basis of the next generation of optical clocks \cite{Ludlow2015}, inertial sensors \cite{Wu2019,Hardman2016,Freier2016,Menoret2018}, and other metrological devices \cite{Pezze2018}, some of which can take advantage of mesoscopic quantum entanglement to enhance their precision \cite{Cox2016,Kohlhaas2015}.  

For many of these systems both the number of atoms in an ultracold sample and the sample density are key parameters that strongly affect their dynamics.  For example, in Bose-Einstein condensates the mean-field energy is determined by the total number of atoms in a condensate, and in Fermi gases the Fermi energy is similarly determined by atom number.  In experiments that probe scattering dynamics the number of atoms scattered per unit time is a function of both the density and total number of atoms \cite{Horvath2017,Thomas2016,Thomas2018}, and in many other studies, ranging from sub-radiance \cite{Ferioli2020} to quantum droplets \cite{Koch2008,Chomaz2019}, the number and/or density of atoms affects phase transitions.  Thus, the precision which can be achieved in such experiments is ultimately dependent on the stability of both the atomic density and number.

Preparing an ultracold atomic cloud with a fixed number of atoms is a task that remains experimentally challenging due to the inherent complexity of the apparatus, which hybridizes several different technologies from optical, microwave, vacuum, electronic, and mechanical engineering \cite{Lewandowski2003,Streed2006}.  The starting point for an ultracold sample is atoms captured in a magneto-optical trap. These atoms are then typically laser cooled in a sub-Doppler scheme and transferred to a magnetic trap, an optical dipole trap, or a combination of both. There is an inevitable stochastic element to this step and when, finally, this trapped sample is evaporatively cooled to the ultracold domain, atom number fluctuations on the percent level are to be expected even for an optimized experimental cycle.  

In principle all inputs to an experiment can be made arbitrarily stable, but in the final accounting external magnetic fields will drift, laser powers will have some noise, and lab temperatures will change \cite{Chilcott2021}.  At some point, rather than attempting to stabilize an experimental input (a laser power, for instance) in order to stabilize the output (the number of atoms), it becomes more practical to stabilize the output directly.  For this to work we need methods for non-destructively measuring the number of atoms and removing excess atoms in real time.  An early example \cite{Sawyer2012} showed that frequency-modulation spectroscopy of the scalar light shift \cite{Lye2003} provided a non-destructive measurement that could allow for the selection of experimental cycles where the initial number of atoms was in an acceptable range.  Subsequently, Gajdacz \ea~\cite{Gajdacz2016} used the vector light shift to demonstrate that atomic samples of $N \sim 10^6$ atoms could be produced with fluctuations below the atomic shot noise limit of $\textrm{Var}(N) = N$.  In later works \cite{Kristensen2019,Christensen2020} the authors used the same technique to probe number fluctuations in the number of Bose-condensed atoms near the critical temperature.  In the latter three experiments an image of the atomic cloud was taken using dark-field Faraday imaging \cite{Gajdacz2013} which allowed the authors to compute a proxy for the number of atoms in the sample.  Excess atoms were removed by applying a series of radio-frequency (RF) pulses that each removed a small fraction of the remaining atoms.  These experiments all used a far off-resonant laser field to minimize inelastically and off-axis scattered photons and hence heating of the atoms.

While imaging provides a wealth of spatial information, the addition of a camera adds significant complications to real-time data acquisition and processing.  Since the total number of atoms in a sample can be related to the integrated signal in the dark-field technique, in principle one can replace the camera with a sufficiently low-noise photodetector. This replacement has three main advantages.  First, readout and processing of a voltage signal is both simpler and faster than readout and processing of camera pixel counts.  Second, the increased speed at which data can be acquired and processed opens up the possibility of stabilizing the number of atoms by an iterative scheme using multiple stages of measurement and feedback rather than the previously-demonstrated single-stage method \cite{Gajdacz2016,Kristensen2019,Christensen2020}.  A multi-stage method can be less sensitive to changes in the environment and the experimental parameters, making it a potentially more robust stabilization scheme.  Multiple stages of feedback can also eliminate the noise floor imposed on the atom number by the stochastic manner in which atoms are removed \cite{Gajdacz2016}.  Finally, photodetectors are significantly smaller, lighter and consume less power than a camera which may be important for mobile or space-based cold-atom devices where size, weight and power (SWaP) are important \cite{Freier2016,Aveline2020}.

In this article, we demonstrate a simplified method for stabilizing the number of atoms in an ultracold atomic sample.  By combining the polarimetric signals resulting from two probe lasers with different frequencies that interact with the atoms and impinge on two low-noise photodetectors, we obtain a measure of the number of atoms that is insensitive to changes in optical power and photodetector gain.  We use the processed signal to inform a multi-stage feedback protocol that iteratively converges on a target signal and demonstrate a reduction in both short and long-term fluctuations in the number of atoms to a relative cycle-to-cycle stability of $0.45\%$ which is limited by temperature fluctuations, photon shot noise, and beam pointing fluctuations.

\section{Implementation}
\label{sec:Implementation}

\begin{figure*}[htb]
	\centering
	\includegraphics[width=\textwidth]{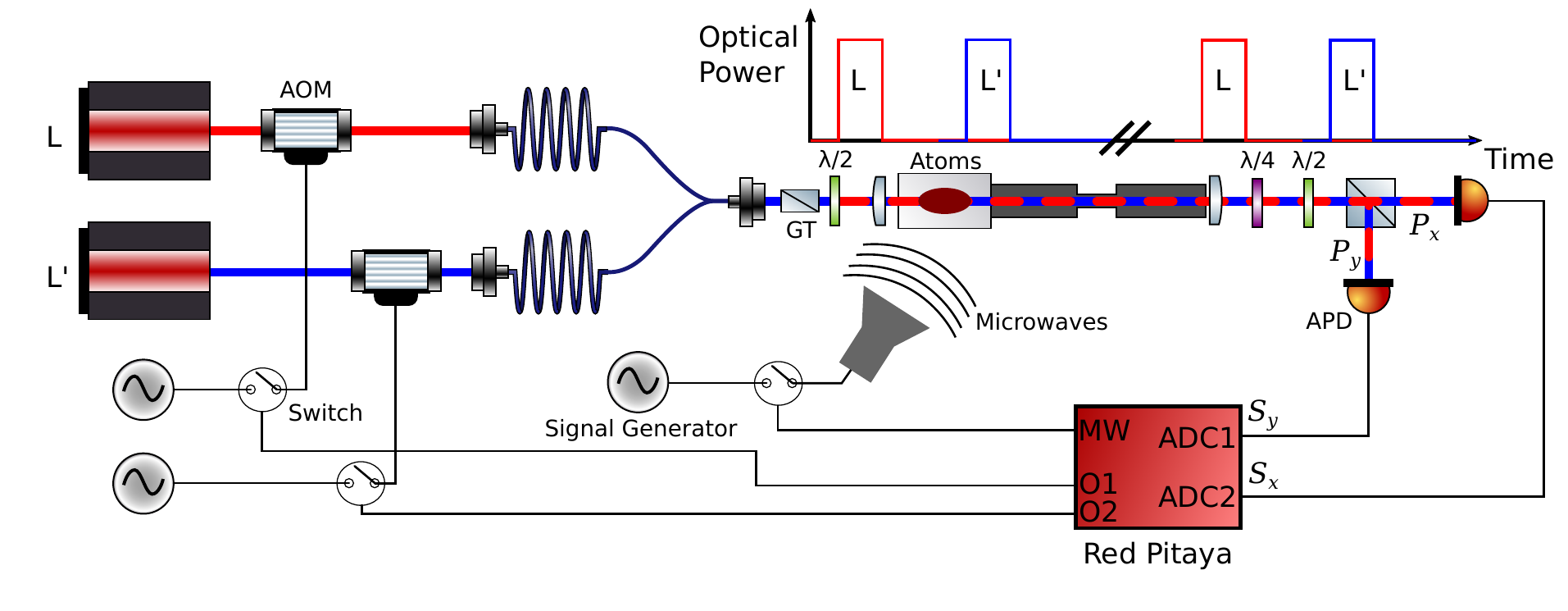}
	\caption{Simplified diagram of the experiment.  Two lasers, routed through two acousto-optic modulators (AOMs), are combined using a fiber beam-combiner/splitter.  The polarization of the two lasers is set by a Glan-Thompson prism (GT), and then rotated to the desired polarization with a half-waveplate.  The combined fields are focused onto the atoms in the science cell using a lens, and the light is collected by another lens after the light has propagated through approximately 1 m of vacuum system including an aperture between high and ultra-high vacuum sections.  A quarter and half-waveplate correct the polarization and set it to $+45^\circ$ before the light passes through a polarizing beam-splitting cube.  The two polarization components impinge on two avalanche photodetectors (APDs), and the two voltage signals are digitized using a Red Pitaya development board which controls the light pulses and the microwave state.}
	\label{fg:setup}
\end{figure*}

Generally, feedback to control a system parameter requires three components: a measurement of the parameter of interest, an actuator that affects the parameter, and a controller that can compute the actuator level from the measurement at speeds faster than the fluctuations that one wishes to remove.  Our system comprises a prolate sample of \Rb{} atoms in the $\ket{F=2,m_F=2}$ hyperfine state confined in a Ioffe-Pritchard (IP) magnetic trap with a radial trapping frequency of $\omega_r = 2\pi\times 160$ Hz and a 10:1 aspect ratio.  We use Faraday rotation \cite{Julsgaard2001,Kaminski2012,Gajdacz2013} of off-resonant, linearly-polarized laser fields propagating along the long-axis of the IP trap to generate a signal that is a proxy for the number of atoms in our sample at a fixed temperature.  

To describe the effect of Faraday rotation, we start by considering an optical field with wavenumber $k$ propagating in the $z$ direction.  We write the laser polarization after the field has passed through the atoms in terms of the left and right-circularly polarized components $\mathbf{\hat{e}}_{\pm}$
\begin{equation}
	\mathbf{E} = E_+ e^{i\phi_+}\mathbf{\hat{e}}_+ + E_- e^{i\phi_-}\mathbf{\hat{e}}_-,
	\label{eq:light-field}
\end{equation}
with amplitudes $E_{\pm}$ and phases $\phi_{\pm}$ which, in general, are functions of the coordinates transverse to the direction of propagation of the light $(x,y)$.  Faraday rotation in atomic systems is caused by a difference in the electric susceptibility for the right and left-circularly polarized components due to differences in the transition dipole matrix elements.  If the field's detuning from resonance $\Delta$ is much larger than the natural linewidth of the optical transition $\Gamma$, we can neglect absorption of the field and write the phase shifts imparted by a gas of transverse density $\rho(x,y)$ as
\begin{equation}
\phi_{\pm}(x,y) = \left(\frac{k}{2\epsilon_0\hbar}\sum_b \frac{|d_{ab}|^2}{\Delta + \Delta_b}\right)\rho(x,y) = \xi_{\pm}\rho(x,y),
\label{eq:phase-shifts}
\end{equation}
where $d_{ab}$ is the dipole matrix element connecting ground state $a$ to excited state $b$, and $\Delta_b$ is the additional detuning associated with excited state $b$ due, for instance, to hyperfine splitting of the excited state levels.  Equations \eqref{eq:light-field} and \eqref{eq:phase-shifts} are valid in the regime where we collect all light in the incident and scattered fields, and where we can neglect diffraction of light during propagation through the sample \cite{Deb2020}.  We use a dual-port measurement scheme as presented in Fig.~\ref{fg:setup} for measuring the rotation angle.  Defining 
\begin{equation}
r_0 = \frac{2|E_+||E_-|}{|E_+|^2+|E_-|^2},
\label{eq:polarization-ratio}
\end{equation}
as the degree of linear polarization, the power in the $x$ (horizontal) and $y$ (vertical) polarization components is
\begin{subequations}
	\begin{align}
		P_x &= \frac{P}{2}\big[1 + r_0 s_a\cos\alpha + r_0c_a\sin\alpha\big]\label{eq:power-x}\\
		P_y &= \frac{P}{2}\big[1 - r_0 s_a\cos\alpha - r_0c_a\sin\alpha\big]\label{eq:power-y},
	\end{align}
\end{subequations}
with total optical power $P$, and 
\begin{subequations}
	\begin{align}
		s_a = \frac{2}{\pi w^2}\int e^{-2(x^2+y^2)/w^2}\sin\phi_a(x,y)dxdy\label{eq:integrated-sin}\\
		c_a = \frac{2}{\pi w^2}\int e^{-2(x^2+y^2)/w^2}\cos\phi_a(x,y)dxdy\label{eq:integrated-cos},
	\end{align}
\end{subequations}
with Gaussian beam waist $w$ and $\phi_+-\phi_- = \pi/2+\alpha+\phi_a$.  Here, $\phi_a(x,y) = (\xi_+-\xi_-)\rho(x,y) = \xi\rho(x,y)$ is the differential phase shift solely due to the atomic sample and $\pi/2+\alpha$ the differential phase shift from all other sources. With these definitions of $r_0$ and $\alpha$, linearly polarized light at $45^\circ$ has $r_0=1$ and $\alpha=0$.  We convert the optical power into voltages using two avalanche photo-detectors (APDs) and then digitize the results using the two analog-to-digital convertors (ADCs) on a Red Pitaya development board to produce signals $S_x$ and $S_y$ which are linearly related to the optical powers $P_x$ and $P_y$ via $S_x = g_x P_x$ and $S_y = g_y P_y$.  The gains $g_x$ and $g_y$ include the APD gains, the ADC conversion gains, and any inefficiencies in coupling the light onto the photodetectors.

If the gains $g_{x}$ and $g_{y}$ are known then one can form linear combinations of the $x$ and $y$ signals $\mathcal{S}_{\pm} = g_yS_x\pm g_xS_y$, and the ratio $\mathcal{R} = \mathcal{S}_-/\mathcal{S}_+$ depends only on the light polarization and the atomic phase shift
\begin{equation}
	\mathcal{R}  = r_0\left(s_a\cos\alpha + c_a\sin\alpha\right).
	\label{eq:ideal-ratio}
\end{equation}
However, small errors in the gain coefficients used in cross-multiplication will lead to small offsets in $\mathcal{R}$ which, if time-dependent, will degrade the precision of the inferred atom number proxy.  We use temperature-compensated APDs (Thorlabs APD430A and APD130A) which have gain stabilities on the order of $1\%$ at ambient temperatures.  Additionally, our laser field propagates along a 2 m path, and although we use lenses to focus the laser onto the APDs, beam-pointing instabilities still lead to changes in the optical power that reaches the detectors which mimic changes in the photodetector gains.  Together, these changes can lead to variations on the order of several percent in our Faraday signal, which, if applied in a feedback loop, ultimately limits the atom number stability to the same level.

Fluctuations in the laser power, laser pointing, and photodetector gains can be circumvented by using two laser frequencies in a differential measurement.  As shown in Fig.~\ref{fg:setup}, we combine two laser fields $L$ and $L'$ using a fiber beam-splitter and interleave pulses from the two lasers.  As long as the time between pulses from $L$ and $L'$ is much shorter than the typical time-scale for variations in photodetector gain and beam-pointing fluctuations we can assume that the effective gains for each laser field are the same.  By choosing the frequencies of $L$ and $L'$ to be on opposite sides of the resonance the difference in phase shifts from the two fields leads to an enhanced signal.  We combine the signals from lasers $L$ and $L'$ to form the ratio
\begin{equation}
R = \frac{S_x S_y' - S_x'S_y}{S_xS_y' + S_x'S_y},
\label{eq:ratio}
\end{equation}
which can be expressed in terms of the physical system as
\begin{equation}
R = \frac{r_0\big[(s_a-s_a')\cos\alpha+(c_a-c_a')\sin\alpha\big]}{1-r_0^2(c_a\sin\alpha+s_a\cos\alpha)(c_a'\sin\alpha+s_a'\cos\alpha)},
\label{eq:ratio-atoms}
\end{equation}
where $s'_a$ and $c'_a$ are the weighted atomic phase differences for laser $L'$, and we assume that the polarizations of $L$ and $L'$ are the same.  $R$ is insensitive to changes in the optical powers of either $L$ or $L'$, drift in the gains of the photodetectors, and, by sharing a common path and spatial mode, $R$ is also insensitive to small changes in the beam alignment onto the APDs.  If we have a Gaussian density profile with $N$ atoms and spatial standard deviation $\kappa = \sqrt{k_B T/m\omega_r^2}$ with temperature $T$ and mass $m$, and we operate in a regime where the phase shifts are small, corresponding to large laser detunings and/or low-density atomic samples, $R$ can be approximated as
\begin{equation}
R \approx \frac{2(\xi - \xi')N}{\pi(w^2+4\kappa^2)}r_0\cos\alpha - \frac{(\xi^2 - \xi'^2)N^2}{4\pi^2\kappa^2(w^2+2\kappa^2)}r_0\sin\alpha.
\label{eq:approximate-ratio}
\end{equation}
Ideally, $\alpha = 0$ and $r_0 = 1$ corresponding to linearly polarized light at $45^\circ$, although small deviations from these values only lead to either a reduced signal strength or a weak second-order dependence on the peak differential phase shift.  We set the polarization at the final polarizing beam-splitter in the absence of atoms by first adjusting the last two waveplates to minimize the amount of light on one APD, and then setting the final half-waveplate such that output of the APD is half of its maximum value.  This method allows us to compensate for any birefringence in the vacuum windows or the dielectric-coated optics, as polarization changes that occur after the atoms can always be reversed.  An important feature of our technique is that the performance of feedback is insensitive to small errors in the polarization as long as those errors are independent of time, since the computation of $R$ from measurements makes no assumption about the polarization of the light.

As seen in Eq.~\eqref{eq:approximate-ratio}, the waist of the laser field is an important parameter for feedback.  When the waist is much smaller than the size of the sample, $R$ is a sensitive probe of the central density of the sample and attains its maximum signal strength.  When $w\gg \kappa$, however, $R$ measures the atom number $N$ directly with minimal dependence on sample size but with a reduced signal strength.  Our probe beam waist of $w \approx 60$ \si{\micro\meter} is approximately the same as the radial size of the atomic sample at the temperature we investigate as a compromise between signal-to-noise and our ability to collect the light at the far end of our vacuum system.  

\begin{figure}[t]
	\centering
	\includegraphics[width=\figwidth]{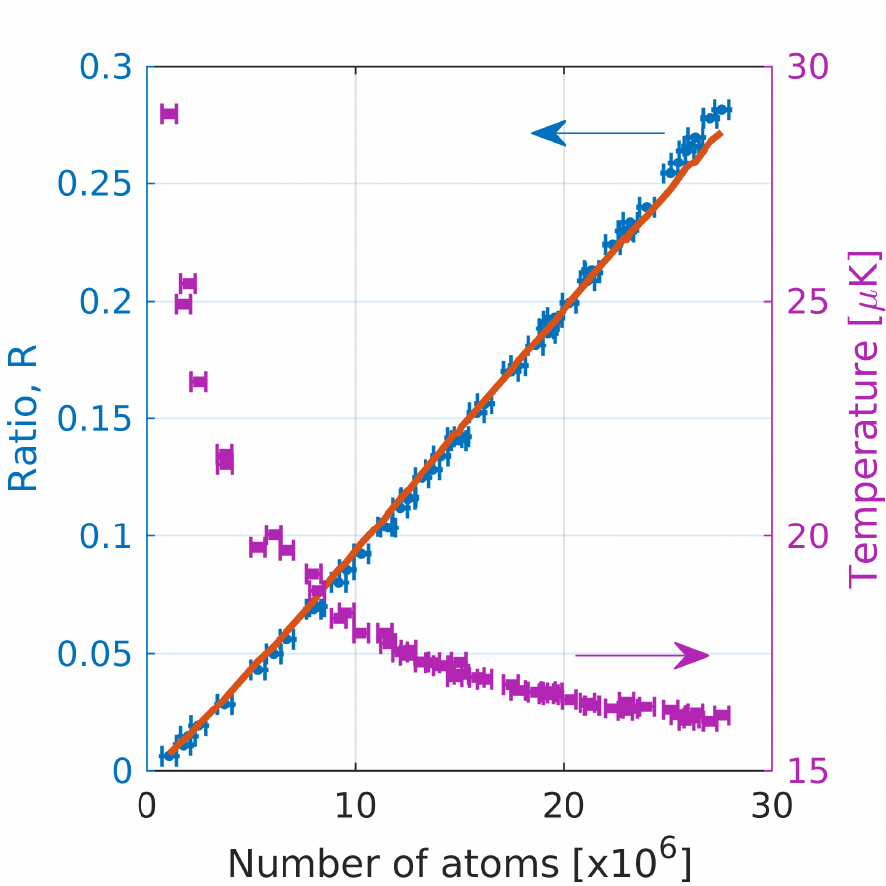}
	\caption{Measured value of $R$ (blue circles, left axis) and sample temperature (purple squares, right axis) as a function of the number of atoms $N$ at a detuning of $-3$ \si{\giga\hertz} for $L$ and $6.8$ GHz for $L'$.  Due to the manner in which the number of atoms is changed, the temperature of the sample also changes.  The solid red line is the prediction from Eq.~\eqref{eq:ratio-atoms} assuming we undercount the number of atoms by $7.3\%$.}
	\label{fg:ratio-vs-number}
\end{figure}

One of our laser fields is sourced from a laser that is offset-locked $-3$ GHz from the \Rb{} $F=2\rightarrow F'=3$ transition ($L$), and the other field is sourced from our MOT repump laser which is locked to the $F=1\rightarrow F'=2$ transition ($L'$).  The power incident on the atoms is $2.5$ \si{\micro\watt} for laser $L$ and $5.0$ \si{\micro\watt} for laser $L'$, but we collect only $\sim$$40\%$ of the light at the final polarizing beam-splitter due to losses on the vacuum chamber windows from adsorbed rubidium.  From these parameters, we expect to pump atoms to other hyperfine states at a fractional rate of $5.2\times 10^{-5}$ $\si{\micro\second^{-1}}$ and to heat the sample at a rate of $26$ $\si{\pico\kelvin\per\micro\second}$.  Figure~\ref{fg:ratio-vs-number} shows how the measured value of $R$ changes when the number of atoms at a fixed trap depth as measured by absorption imaging is varied by changing the number of atoms loaded into our IP trap.  Due to changes in evaporative cooling efficiency and thermalization rates with the number of atoms loaded into the IP trap, the temperature of the sample varies with the number of atoms from $29$ \si{\micro\kelvin} for the lowest number of atoms to $16.2$ \si{\micro\kelvin} for the highest number of atoms.  We get good agreement with the theoretical prediction of Eq.~\eqref{eq:ratio-atoms} if we assume that we undercount the number of atoms by $7.3\%$, which can arise from a small change in the direction of the background magnetic field experienced by the atoms during imaging compared to our expectation.

\begin{figure}[tb]
	\centering
	\includegraphics[width=\figwidth]{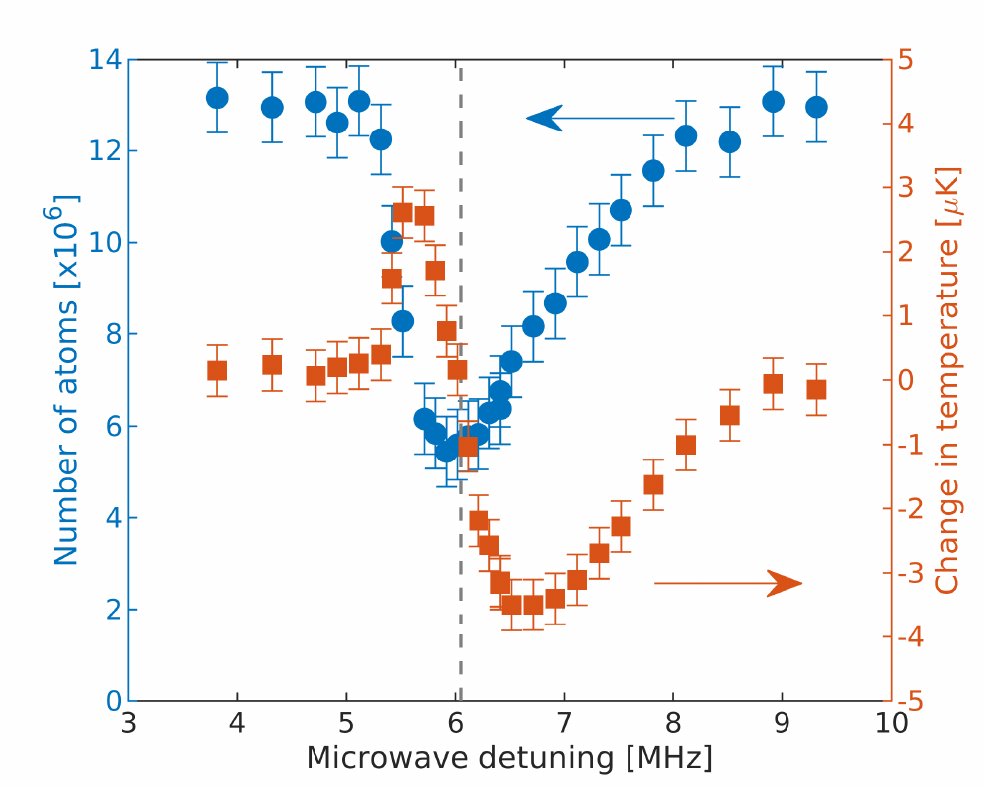}
	\caption{Effect of microwave pulses on a sample of atoms at an initial temperature of $16$ \si{\micro\kelvin} for $1000$ microwave pulses that are $25$ \si{\micro\second} wide and spaced by $50$ \si{\micro\second}.  Detuning is relative to the microwave frequency at zero magnetic field.  Left axis (blue, circles) shows the remaining number of atoms while the right axis (red, squares) shows the change in temperature as a function of microwave frequency.  The gray dashed line shows the approximate microwave frequency at which the change in temperature is zero.}
	\label{fg:microwave-calibration}
\end{figure}

For our actuator, we use microwave pulses resonant with the $\ket{2,2}\rightarrow\ket{1,1}$ transition at $\sim$$6.8$~GHz.  We follow previous work \cite{Gajdacz2016,Kristensen2019} and use many short pulses to remove atoms from the sample by transferring small fractions to the anti-trapped $\ket{1,1}$ state where atoms are expelled from the trap.  We use $25$ \si{\micro\second} long pulses which each transfer approximately $f \approx 10^{-3}$ of the atoms in the $\ket{2,2}$ state to the $\ket{1,1}$ state, and the microwave pulses are spaced by $50$ \si{\micro\second}.  For sufficiently small transfer fraction per pulse, we can approximate the number of pulses needed to remove a given fraction of atoms as a linear function of $f^{-1}$.  This choice of many short pulses, as opposed to a single longer pulse, simplifies the calculation of the actuator value and avoids transferring a large fraction of atoms to the $\ket{1,1}$ state where the $L'$ laser is close to resonance and thus would be more strongly affected by the anti-trapped atoms than the trapped atoms.  As our sample is magnetically trapped, and the natural linewidth of the microwave transition is negligible, the microwave field only addresses atoms in a narrow energy range whose width is defined by the temporal width and Rabi frequency of the pulse.  Therefore, the number of atoms removed as a function of microwave frequency will reflect the atoms' potential energy distribution as shown in Fig.~\ref{fg:microwave-calibration}.  An important consequence of this energy selectivity is that the microwave pulses will heat (cool) the sample when the microwave frequency is set too low (high).  Theoretically, the microwave frequency should be set to be resonant with the mean energy of the sample \cite{Kristensen2019}; in practice, we experimentally determine the optimum frequency by measuring the change in temperature of the sample as a function of microwave frequency as in Fig.~\ref{fg:microwave-calibration}.

\begin{figure}[tb]
	\centering
	\includegraphics[width=\figwidth]{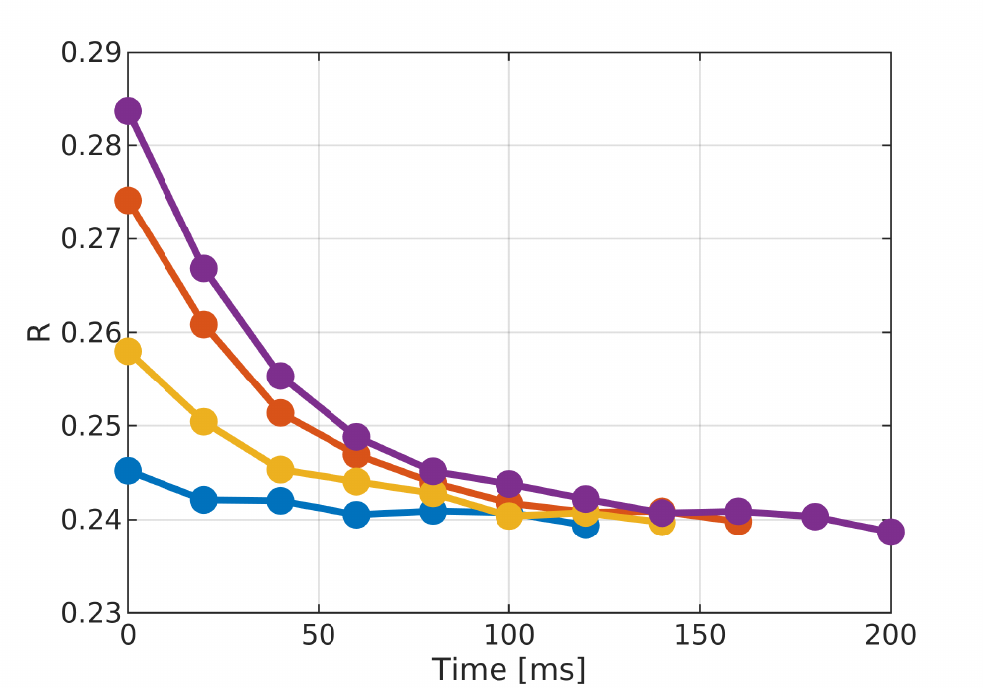}
	\caption{Example data showing the decay of the value of $R$ during feedback.  Each set is a different run of the experiment with different initial numbers of atoms at  fixed temperatures.  The target value was set to $R_{\rm target} = 0.24$ with a tolerance of $10^{-3}$.  The value of $b$ for these curves is $b\approx0.4$.}
	\label{fg:ratio-decay}
\end{figure}

To unite our measurement with our actuator and close the feedback loop, we use programmable logic on board a 14-bit Red Pitaya (RP) development board.  Our feedback protocol proceeds as follows.  First, the RP pulses on $L$ then $L'$ for a programmable duration with the $L'$ pulse delayed relative to the $L$ pulse, and the resulting APD voltages are captured by the two ADCs on the RP.  The signal values $S_{x,y}$ and $S'_{x,y}$ are calculated as the difference between the mean APD voltages when the pulses are on and when the pulses are off to account for any offset voltages on the ADCs.  The ratio $R_{j-1}$ is then calculated from the integrated signals as in Eq.~\eqref{eq:ratio} to 16 bits of precision.  $R_{j-1}$ is compared with a target value, $R_{\rm target}$, and we calculate the number of microwave pulses $n_j$ to apply for the current feedback round $j$ as
\begin{equation}
	n_j = \left\lfloor bf^{-1}\left(\frac{R_{j-1}-R_{\rm target}}{R_{j-1}}\right)\right\rfloor,
	\label{eq:number-of-pulses}
\end{equation}
with $b$ the fraction of excess atoms to remove in a given round $j$ and $R_0$ the initial value of $R$.  After $j$ rounds of feedback, the measured ratio will be 
\begin{equation}
	R_j = R_{\rm target}+(R_{0}-R_{\rm target})(1-b)^j.
	\label{eq:iterative-feedback}
\end{equation}
The feedback protocol described by Ref.~\citenum{Gajdacz2016} effectively implements Eq.~\eqref{eq:iterative-feedback} with $b=1$ where all the excess atoms are removed after a single measurement.  If, however, the fraction of atoms removed with each microwave pulse changes or is not well-known, then a single round of feedback will not necessarily produce a sample with the target number of atoms.  Additionally, the stochastic nature of removing atoms leads to an uncertainty in the final number of atoms that increases with the number of atoms removed \cite{Hume2013,Gajdacz2016}.  Both of these effects result in the final fluctuations in the number of atoms depending on the number of atoms removed relative to the target value in the last feedback round.  Therefore, we use $b=0.25$ to $b=0.5$, and we terminate feedback when $R_j < (1+\textrm{tol})R_{\rm target}$ where $\textrm{tol}$ is a fractional tolerance value that we set to $1\times 10^{-3}$.  Example decay curves of $R$ are shown in Fig.~\ref{fg:ratio-decay} for $b\approx 0.4$, corresponding to $f\approx 1.5\times 10^{-3}$, and $R_{\rm target} = 0.24$. The number of atoms to remove in the last round relative to the target is then bounded from above by
\begin{equation}
\frac{R_{j-1} - R_{\rm target}}{R_{\rm target}} \leq \frac{\textrm{tol}}{1-b},
\label{eq:upper-bound-fraction-removed}
\end{equation}
which is $1.7\times 10^{-3}$ for the parameters given above.  Relative fluctuations in the final number of atoms caused by relative fluctuations in $f$ are further reduced by a factor of $b$, resulting in a suppression factor of $6.7\times 10^{-4}$.  Fluctuations in the final number caused by the stochastic removal of atoms are reduced below the shot noise level by the square root of this factor, or $2.6\times 10^{-2}$, regardless of the initial number of atoms.

\section{Results}
\label{sec:results}
\begin{figure}[tb]
	\centering
	\includegraphics[width=\figwidth]{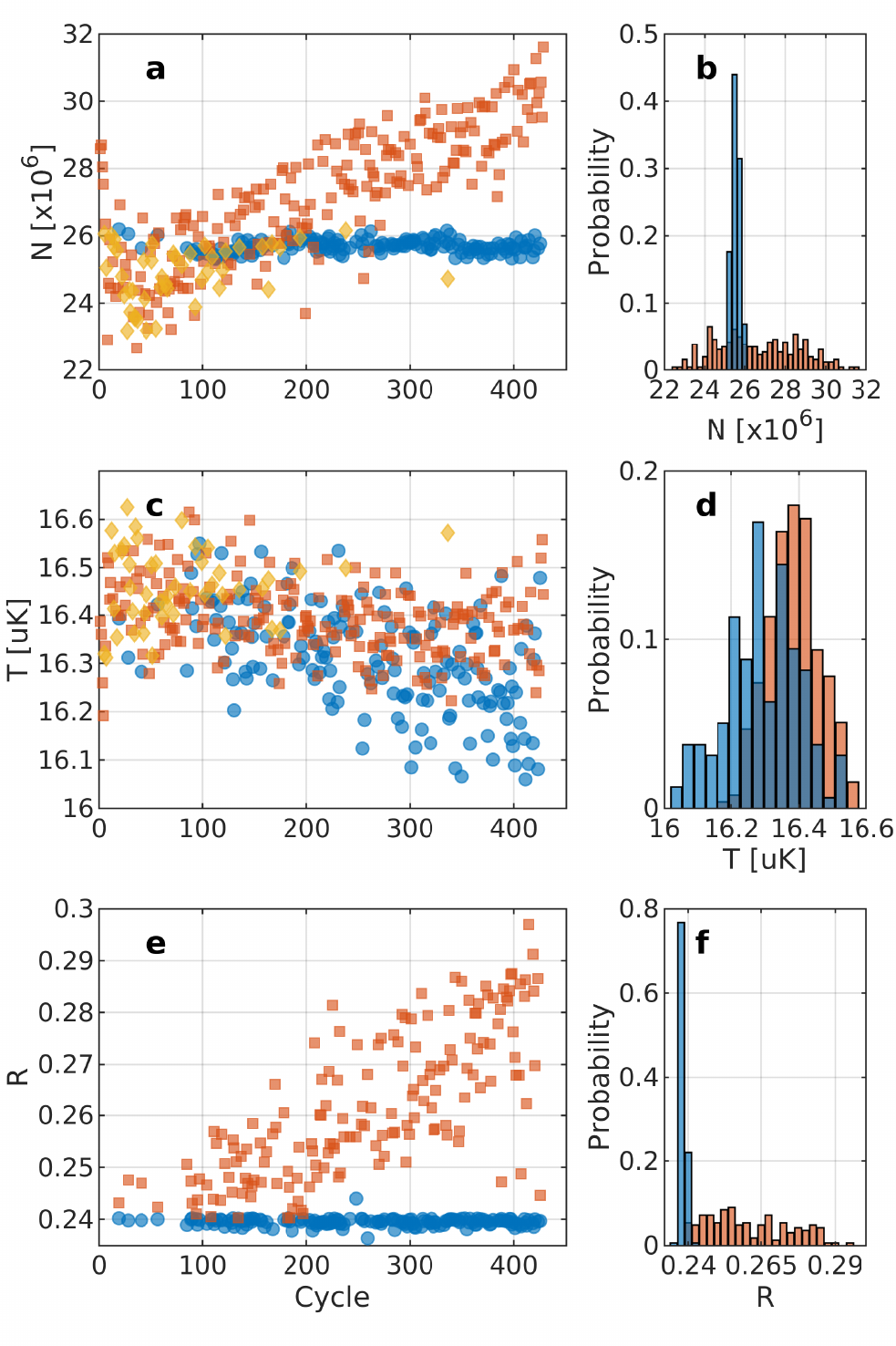}
	\caption{Effect of feedback on a sample of atoms.  Red squares and bars indicate cycles when feedback was disabled, blue circles and bars indicate when feedback was applied, and yellow diamonds indicate when feedback was enabled but $R_0 < R_{\rm target}$ and feedback was not applied.  \textbf{a,b} Number of atoms in the sample as measured by absorption imaging. \textbf{c,d} Temperature of the sample.  \textbf{e,f} Initial and final values of $R$ when feedback was applied.  Blue circles and bars indicate the final value of $R$, and red squares and bars indicate the initial value $R_0$.}
	\label{fg:time-series}
\end{figure}
Figure~\ref{fg:time-series} shows the effect of our iterative feedback scheme on a sample of \Rb{} atoms held in our IP trap at a temperature of \SI{16.3}{\micro\kelvin}, where we have set $R_{\rm target} = 0.24$ corresponding to $N_{\rm target} = 25.7\times 10^6$ atoms.  Each round of feedback comprises a measurement of $R$ using pulses from $L$ and $L'$ that are $50$ \si{\micro\second} long with the pulses from $L'$ delayed by \SI{1.5}{\milli\second}, followed by application of a variable number of microwave pulses.  Each pair of pulses, which define the start of a feedback round, are separated in time by 20 ms which is chosen to allow enough time for all possible microwave pulses to be applied.  The number of microwave pulses is calculated by the Red Pitaya using Eq.~\eqref{eq:number-of-pulses} with $bf^{-1}=250$ defining the maximum number of microwave pulses that can be applied.  The exact value for the maximum number of microwave pulses is unimportant, except that a larger value leads to fewer feedback rounds at the cost of greater sensitivity to fluctuations in $f$.  

We measure the number of atoms and the temperature at the end of a cycle using standard absorption imaging resonant with the $F=2\rightarrow F'=3$ transition along one of the radial axes after releasing the atoms from the IP trap and waiting 15 ms.  This serves as an important independent measure of the number of atoms, as feedback in general suppresses fluctuations in the measurement ($R$) but not necessarily in the system state ($N$) \cite{Bechhoefer2005}.  Our absorption imaging light is turned on for $250$ \si{\micro\second} at an intensity of $670$ \si{\micro\watt\per\centi\meter\squared}.  The quantum efficiency of the camera sensor at our imaging wavelength is $0.8$, and the photon shot noise dominates over the readout noise with our effective pixel size of $\SI{11}{\micro\meter}\times\SI{11}{\micro\meter}$.  From these parameters, we attribute relative fluctuations on the number of atoms of $0.014\%$ from photon shot noise.  A much larger contribution comes from inhomogeneity in the imaging fields, such as fringes caused by the relatively long separation between pairs of images of $18$ \si{\milli\second}, which we estimate from Monte Carlo simulations of the fitting process and images to be $0.4\%$.

The data in Fig.~\ref{fg:time-series} comprise 425 experimental cycles over the course of 10 hours where we randomly choose to enable or disable feedback with $50\%$ probability.  When feedback is disabled we do not apply any optical pulses in order to measure the free-running distribution of atom numbers and temperatures.  As can be seen in Figs.~\ref{fg:time-series}(a)-(d), our experiment shows significant cycle-to-cycle variations in the number of atoms in addition to long-term drift over the course of a day while the temperature remains nearly constant.  We set $R_{\rm target}$ such that the desired number of atoms is close to the initial number of atoms at the start of the day.  Early in the series, feedback engages sporadically as $R_0 < R_{\rm target}$; however, these cases are easily detected in post-processing as only one feedback round is attempted.  As a result, in experiments where the free-running number of atoms shows no drift the number can be stabilized to the mean, free-running value with only a $50\%$ reduction in the experimental duty cycle. This will be advantageous when the variance in the free-running number of atoms is more than twice the variance in the stabilized number of atoms as the decreased variance will more than compensate for the increased cycle time.

In the latter half of the time series of Fig.~\ref{fg:time-series} the free-running number of atoms almost always exceeds the target value, and thus feedback is almost always applied.  The result is that the cycle-to-cycle fluctuations and drift in the number of atoms are significantly reduced.  Figure~\ref{fg:time-series}(c) and Fig.~\ref{fg:time-series}(d) show that the temperature of the sample under feedback tends to be slightly lower than the temperature absent feedback.  This change in temperature is caused by setting the microwave frequency slightly too high relative to the mean energy of the atoms.  The change in temperature is equivalent to $\approx -2$ \si{\nano\kelvin\per\textrm{pulse}}, and this change is noticeable in the latter part of the time series because more microwave pulses are required to remove enough atoms to reach the target value.  This effect dominates over the expected heating rate from the optical pulses which is approximately $1.3$ \si{\nano\kelvin} per feedback round.

\begin{figure}[tb]
	\centering
	\includegraphics[width=\figwidth]{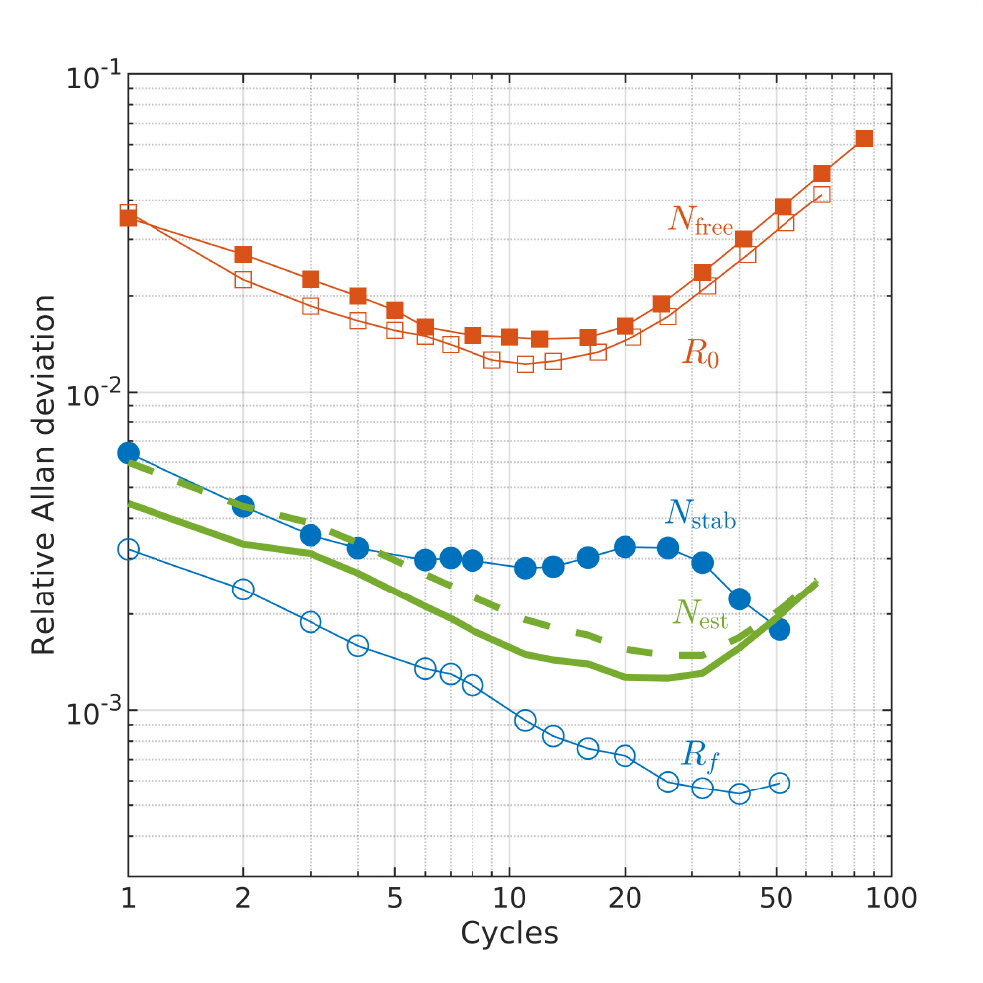}
	\caption{Relative Allan deviations of the number of atoms $N$ and the ratio $R$.  The Allan deviations for the number of atoms are indicated by filled markers, where blue circles are the Allan deviation with feedback applied ($N_{\rm stab}$) and the red squares are without feedback ($N_{\rm free}$).  The Allan deviations for $R$ when feedback is applied are indicated by open markers, where blue circles are for the final value of $R$ when feedback terminates ($R_f$), and red squares are the Allan deviation for initial values ($R_0$).  The solid green line is relative Allan deviation from the estimated number of atoms $N_{\rm est}$ from $R_f$, the measured temperature, and Eq.~\eqref{eq:approximate-ratio}, and the dashed green line is the sum-in-quadrature of the Allan deviation of $N_{\rm est}$ and the estimated imaging noise.}
	\label{fg:allan-deviation}
\end{figure} 

To quantify the variation in the number of atoms we use the overlapped estimator of the Allan variance \cite{IEEEStandard}
\begin{align}
\sigma_x^2(\tau) &= \frac{1}{2(M-2\tau)\tau^2}\times\notag\\
&\sum_{k=0}^{M-2\tau-1}\left[\sum_{j=\tau+k}^{2\tau+k-1}x_j - \sum_{j=k}^{\tau+k-1}x_j\right]^2,
\label{eq:allan-variance}
\end{align}
where $M$ is the number of samples in a data set $\{x_j\}$ and $\tau$ is the offset at which to calculate the Allan variance $\sigma^2(\tau)$.  We plot the relative Allan deviation $\sqrt{\sigma_x^2(\tau)}/\langle x\rangle$ in Fig.~\ref{fg:allan-deviation} for the number of atoms measured with and without feedback (filled markers), labelled $N_{\rm free}$ and $N_{\rm stab}$, respectively, and the initial ($R_0$) and final ($R_f$) values of $R$ when feedback is applied (open markers).  Cycle-to-cycle variations of $N_{\rm free}$ and $R_0$ when feedback is disabled are about $3.5\%$, and both show significant drift over the course of a day.  In contrast, the cycle-to-cycle variations in the controlled values of $N_{\rm stab}$ and $R_f$ show relative fluctuations of $0.65\%$ and $0.32\%$, respectively.  To assess the ``true'' performance of our method, we estimate the number of atoms from $R_f$ and our measurements of the temperature $T$ using Eq.~\eqref{eq:approximate-ratio}, and we plot the relative Allan deviation of this quantity as $N_{\rm est}$ in Fig.~\ref{fg:allan-deviation} as the solid green line.  Cycle-to-cycle variations in $N_{\rm est}$ amount to $0.45\%$, and when added in quadrature with the estimated uncertainty from absorption imaging, agree well with the short-term measured variations in $N_{\rm stab}$ as seen by the dashed green line in Fig.~\ref{fg:allan-deviation}.  In the longer term there is more variation in $N_{\rm stab}$ than in $N_{\rm est}$, and this is likely due to drifts in the orientation of the background magnetic field which would affect the measurement of the number of atoms but not the temperature.  For long averaging times, $N_{\rm est}$ starts to show some drift as evidenced by the upward trajectory of the relative Allan deviation in Fig.~\ref{fg:allan-deviation}.  This is caused by the change in temperature when feedback is enabled as seen in Fig.~\ref{fg:time-series}c, and could be reduced by a more careful calibration of the microwave frequency or use of a larger beam size at the expense of lower signal-to-noise.

Our cycle-to-cycle fluctuations in $R$ limit the final stability of the atom number since feedback cannot stabilize a parameter to a higher precision than one can measure that parameter.  Based on our applied laser powers and pulse durations as well as our detectors' noise-equivalent powers ($\SI{140}{\pico\watt\hertz^{-1/2}}$) and the noise floor of the Red Pitaya's ADCs ($\SI{50}{\nano\volt\hertz^{-1/2}}$), our measurement is dominated by photon shot noise \cite{ExcessNoise}, and this imposes a limit on the cycle-to-cycle stability of $0.15\%$.  Stochastic loss of atoms to other hyperfine states due to optical pumping can lead to additional noise \cite{Hume2013}, but we have estimated this to contribute relative fluctuations of only $2\times 10^{-5}$ based on aforementioned loss rates and pulse durations.  Additional pulses applied after feedback has terminated confirm that our final stability is due to fluctuations from pulse-to-pulse, not from cycle-to-cycle.  We attribute the excess noise on our measurement of $R$ to beam-pointing noise arising from two different processes.  First, beam pointing noise onto the atoms causes changes in $R$ as the overlap between the sample and the laser fields changes.  Second, the sample acts as a lens \cite{Deb2020} with different focal lengths for the different circular polarization components, and a small change in the beam position on the atoms will map to different final beam positions for the different polarizations.  Effectively, this leads to pulse-to-pulse changes in $r_0$ and hence $R$.  Reducing beam-pointing noise would require significant re-design of our apparatus, but other apparatuses need not suffer the same problem and may be able to reach the limit imposed by photon shot noise.

\section{Conclusion}
\label{sec:conclusion}

In this article we have demonstrated a simple method for reducing the fluctuations in the number of atoms in an ultracold sample.  We use a differential measurement of off-resonant Faraday rotation of two laser fields as they propagate through the sample to compute a proxy for the number of atoms that is independent of changes in laser power and photodetector gain and is only weakly sensitive to alignment.  For sufficiently large detunings, this proxy value is linear in the number of atoms for a fixed temperature.  By using an iterative feedback scheme, we reduce the need for pre-calibration of our control system which makes it more robust to environmental perturbations.  We independently measure fluctuations in the number of atoms using absorption imaging, and we demonstrate a reduction in the cycle-to-cycle number fluctuations down to $0.45\%$ which is limited by temperature fluctuations, photon shot noise, and beam pointing noise.  

Apparatuses with better optical design will likely be able to eliminate beam pointing as a source of noise, leaving temperature fluctuations and photon shot noise as the leading contributions.  Sensitivity to temperature changes can be reduced by using a beam size that is larger than the size of the sample at a cost of reduced signal-to-noise.  Alternatively, feedback could be applied to atoms trapped in an optical dipole trap, where microwave fields are resonant with the entire sample and trapping parameters are more easily controlled, although the actuator would then require a dissipative element such as a resonant optical field.  Supposing that these two sources of noise can be eliminated, the effect of photon shot noise can be strongly reduced by using higher optical powers and/or longer pulse durations.  A straightforward calculation shows that with the optical powers used in this study one can stabilize $26\times 10^6$ atoms to a factor of two below the atomic shot noise level with pulses that are $4$ \si{\milli\second} long \cite{Hume2013}.  A two step feedback method \cite{Gajdacz2016}, where one removes most of the atoms in the first step and then removes only a small fraction in the second step, may be required in this situation to minimize heating of the sample and loss to other hyperfine states.

Although our reported cycle-to-cycle fluctuations in our feedback measurement are a factor of $\sim$$5$ larger than the state-of-the-art \cite{Gajdacz2016,Kristensen2019}, our control system has a number of advantages.  It is both simpler and less costly in terms of equipment and data processing, and it is also faster which allows for an iterative scheme to be used that is more robust to calibration errors.  In applications requiring portable atom-based devices, our system has signficant advantages in terms of SWaP.  Finally, by using microwave pulses instead of radio-frequency pulses our method is easily adapted to feedback control of the number of atoms in dual species experiments.

Upon preparing this article, we became aware of recent related work using an optical cavity for dispersive real time tracking of evaporative cooling with cavity-assisted feedback stabilization of atom number as a future goal \cite{Zeiher2020}.

\section*{Acknowledgements}
This work was funded by the New Zealand Tertiary Education Commission through the Dodd-Walls Centre for Photonic and Quantum Technologies.

\section*{Data Availability Statement}
Project files, hardware definition language (HDL) code for the FPGA architecture, and software (including documentation) for implementing feedback control can be found on GitHub \cite{GitHub}.


\begin{thebibliography}{46}%
\makeatletter
\providecommand \@ifxundefined [1]{%
 \@ifx{#1\undefined}
}%
\providecommand \@ifnum [1]{%
 \ifnum #1\expandafter \@firstoftwo
 \else \expandafter \@secondoftwo
 \fi
}%
\providecommand \@ifx [1]{%
 \ifx #1\expandafter \@firstoftwo
 \else \expandafter \@secondoftwo
 \fi
}%
\providecommand \natexlab [1]{#1}%
\providecommand \enquote  [1]{``#1''}%
\providecommand \bibnamefont  [1]{#1}%
\providecommand \bibfnamefont [1]{#1}%
\providecommand \citenamefont [1]{#1}%
\providecommand \href@noop [0]{\@secondoftwo}%
\providecommand \href [0]{\begingroup \@sanitize@url \@href}%
\providecommand \@href[1]{\@@startlink{#1}\@@href}%
\providecommand \@@href[1]{\endgroup#1\@@endlink}%
\providecommand \@sanitize@url [0]{\catcode `\\12\catcode `\$12\catcode
  `\&12\catcode `\#12\catcode `\^12\catcode `\_12\catcode `\%12\relax}%
\providecommand \@@startlink[1]{}%
\providecommand \@@endlink[0]{}%
\providecommand \url  [0]{\begingroup\@sanitize@url \@url }%
\providecommand \@url [1]{\endgroup\@href {#1}{\urlprefix }}%
\providecommand \urlprefix  [0]{URL }%
\providecommand \Eprint [0]{\href }%
\providecommand \doibase [0]{https://doi.org/}%
\providecommand \selectlanguage [0]{\@gobble}%
\providecommand \bibinfo  [0]{\@secondoftwo}%
\providecommand \bibfield  [0]{\@secondoftwo}%
\providecommand \translation [1]{[#1]}%
\providecommand \BibitemOpen [0]{}%
\providecommand \bibitemStop [0]{}%
\providecommand \bibitemNoStop [0]{.\EOS\space}%
\providecommand \EOS [0]{\spacefactor3000\relax}%
\providecommand \BibitemShut  [1]{\csname bibitem#1\endcsname}%
\let\auto@bib@innerbib\@empty
\bibitem [{\citenamefont {Burger}\ \emph {et~al.}(1999)\citenamefont {Burger},
  \citenamefont {Bongs}, \citenamefont {Dettmer}, \citenamefont {Ertmer},
  \citenamefont {Sengstock}, \citenamefont {Sanpera}, \citenamefont
  {Shlyapnikov},\ and\ \citenamefont {Lewenstein}}]{Burger1999}%
  \BibitemOpen
  \bibfield  {author} {\bibinfo {author} {\bibfnamefont {S.}~\bibnamefont
  {Burger}}, \bibinfo {author} {\bibfnamefont {K.}~\bibnamefont {Bongs}},
  \bibinfo {author} {\bibfnamefont {S.}~\bibnamefont {Dettmer}}, \bibinfo
  {author} {\bibfnamefont {W.}~\bibnamefont {Ertmer}}, \bibinfo {author}
  {\bibfnamefont {K.}~\bibnamefont {Sengstock}}, \bibinfo {author}
  {\bibfnamefont {A.}~\bibnamefont {Sanpera}}, \bibinfo {author} {\bibfnamefont
  {G.~V.}\ \bibnamefont {Shlyapnikov}},\ and\ \bibinfo {author} {\bibfnamefont
  {M.}~\bibnamefont {Lewenstein}},\ }\bibfield  {title} {\bibinfo {title} {Dark
  {Solitons} in {Bose}-{Einstein} {Condensates}},\ }\href
  {https://doi.org/10.1103/PhysRevLett.83.5198} {\bibfield  {journal} {\bibinfo
   {journal} {Phys. Rev. Lett.}\ }\textbf {\bibinfo {volume} {83}},\ \bibinfo
  {pages} {5198} (\bibinfo {year} {1999})}\BibitemShut {NoStop}%
\bibitem [{\citenamefont {Everitt}\ \emph {et~al.}(2017)\citenamefont
  {Everitt}, \citenamefont {Sooriyabandara}, \citenamefont {Guasoni},
  \citenamefont {Wigley}, \citenamefont {Wei}, \citenamefont {McDonald},
  \citenamefont {Hardman}, \citenamefont {Manju}, \citenamefont {Close},
  \citenamefont {Kuhn}, \citenamefont {Szigeti}, \citenamefont {Kivshar},\ and\
  \citenamefont {Robins}}]{Everitt2017}%
  \BibitemOpen
  \bibfield  {author} {\bibinfo {author} {\bibfnamefont {P.~J.}\ \bibnamefont
  {Everitt}}, \bibinfo {author} {\bibfnamefont {M.~A.}\ \bibnamefont
  {Sooriyabandara}}, \bibinfo {author} {\bibfnamefont {M.}~\bibnamefont
  {Guasoni}}, \bibinfo {author} {\bibfnamefont {P.~B.}\ \bibnamefont {Wigley}},
  \bibinfo {author} {\bibfnamefont {C.~H.}\ \bibnamefont {Wei}}, \bibinfo
  {author} {\bibfnamefont {G.~D.}\ \bibnamefont {McDonald}}, \bibinfo {author}
  {\bibfnamefont {K.~S.}\ \bibnamefont {Hardman}}, \bibinfo {author}
  {\bibfnamefont {P.}~\bibnamefont {Manju}}, \bibinfo {author} {\bibfnamefont
  {J.~D.}\ \bibnamefont {Close}}, \bibinfo {author} {\bibfnamefont {C.~C.~N.}\
  \bibnamefont {Kuhn}}, \bibinfo {author} {\bibfnamefont {S.~S.}\ \bibnamefont
  {Szigeti}}, \bibinfo {author} {\bibfnamefont {Y.~S.}\ \bibnamefont
  {Kivshar}},\ and\ \bibinfo {author} {\bibfnamefont {N.~P.}\ \bibnamefont
  {Robins}},\ }\bibfield  {title} {\bibinfo {title} {Observation of a
  modulational instability in {Bose}-{Einstein} condensates},\ }\href
  {https://doi.org/10.1103/PhysRevA.96.041601} {\bibfield  {journal} {\bibinfo
  {journal} {Phys. Rev. A}\ }\textbf {\bibinfo {volume} {96}},\ \bibinfo
  {pages} {041601(R)} (\bibinfo {year} {2017})}\BibitemShut {NoStop}%
\bibitem [{\citenamefont {Papp}\ \emph {et~al.}(2008)\citenamefont {Papp},
  \citenamefont {Pino},\ and\ \citenamefont {Wieman}}]{Papp2008}%
  \BibitemOpen
  \bibfield  {author} {\bibinfo {author} {\bibfnamefont {S.~B.}\ \bibnamefont
  {Papp}}, \bibinfo {author} {\bibfnamefont {J.~M.}\ \bibnamefont {Pino}},\
  and\ \bibinfo {author} {\bibfnamefont {C.~E.}\ \bibnamefont {Wieman}},\
  }\bibfield  {title} {\bibinfo {title} {Tunable {Miscibility} in a
  {Dual}-{Species} {Bose}-{Einstein} {Condensate}},\ }\href
  {https://doi.org/10.1103/PhysRevLett.101.040402} {\bibfield  {journal}
  {\bibinfo  {journal} {Phys. Rev. Lett.}\ }\textbf {\bibinfo {volume} {101}},\
  \bibinfo {pages} {040402} (\bibinfo {year} {2008})}\BibitemShut {NoStop}%
\bibitem [{\citenamefont {Ospelkaus}\ \emph {et~al.}(2006)\citenamefont
  {Ospelkaus}, \citenamefont {Ospelkaus}, \citenamefont {Humbert},
  \citenamefont {Sengstock},\ and\ \citenamefont {Bongs}}]{Ospelkaus2006}%
  \BibitemOpen
  \bibfield  {author} {\bibinfo {author} {\bibfnamefont {S.}~\bibnamefont
  {Ospelkaus}}, \bibinfo {author} {\bibfnamefont {C.}~\bibnamefont
  {Ospelkaus}}, \bibinfo {author} {\bibfnamefont {L.}~\bibnamefont {Humbert}},
  \bibinfo {author} {\bibfnamefont {K.}~\bibnamefont {Sengstock}},\ and\
  \bibinfo {author} {\bibfnamefont {K.}~\bibnamefont {Bongs}},\ }\bibfield
  {title} {\bibinfo {title} {Tuning of {Heteronuclear} {Interactions} in a
  {Degenerate} {Fermi}-{Bose} {Mixture}},\ }\href
  {https://doi.org/10.1103/PhysRevLett.97.120403} {\bibfield  {journal}
  {\bibinfo  {journal} {Phys. Rev. Lett.}\ }\textbf {\bibinfo {volume} {97}},\
  \bibinfo {pages} {120403} (\bibinfo {year} {2006})}\BibitemShut {NoStop}%
\bibitem [{\citenamefont {Ferioli}\ \emph {et~al.}(2021)\citenamefont
  {Ferioli}, \citenamefont {Glicenstein}, \citenamefont {Henriet},
  \citenamefont {Ferrier-Barbut},\ and\ \citenamefont
  {Browaeys}}]{Ferioli2020}%
  \BibitemOpen
  \bibfield  {author} {\bibinfo {author} {\bibfnamefont {G.}~\bibnamefont
  {Ferioli}}, \bibinfo {author} {\bibfnamefont {A.}~\bibnamefont
  {Glicenstein}}, \bibinfo {author} {\bibfnamefont {L.}~\bibnamefont
  {Henriet}}, \bibinfo {author} {\bibfnamefont {I.}~\bibnamefont
  {Ferrier-Barbut}},\ and\ \bibinfo {author} {\bibfnamefont {A.}~\bibnamefont
  {Browaeys}},\ }\bibfield  {title} {\bibinfo {title} {Storage and {Release} of
  {Subradiant} {Excitations} in a {Dense} {Atomic} {Cloud}},\ }\href
  {https://doi.org/10.1103/PhysRevX.11.021031} {\bibfield  {journal} {\bibinfo
  {journal} {Phys. Rev. X}\ }\textbf {\bibinfo {volume} {11}},\ \bibinfo
  {pages} {021031} (\bibinfo {year} {2021})},\ \bibinfo {note} {publisher:
  American Physical Society}\BibitemShut {NoStop}%
\bibitem [{\citenamefont {Inouye}\ \emph {et~al.}(1999)\citenamefont {Inouye},
  \citenamefont {Chikkatur}, \citenamefont {Stamper-Kurn}, \citenamefont
  {Stenger}, \citenamefont {Pritchard},\ and\ \citenamefont
  {Ketterle}}]{Inouye1999}%
  \BibitemOpen
  \bibfield  {author} {\bibinfo {author} {\bibfnamefont {S.}~\bibnamefont
  {Inouye}}, \bibinfo {author} {\bibfnamefont {A.~P.}\ \bibnamefont
  {Chikkatur}}, \bibinfo {author} {\bibfnamefont {D.~M.}\ \bibnamefont
  {Stamper-Kurn}}, \bibinfo {author} {\bibfnamefont {J.}~\bibnamefont
  {Stenger}}, \bibinfo {author} {\bibfnamefont {D.~E.}\ \bibnamefont
  {Pritchard}},\ and\ \bibinfo {author} {\bibfnamefont {W.}~\bibnamefont
  {Ketterle}},\ }\bibfield  {title} {\bibinfo {title} {Superradiant {Rayleigh}
  {Scattering} from a {Bose}-{Einstein} {Condensate}},\ }\href
  {https://doi.org/10.1126/science.285.5427.571} {\bibfield  {journal}
  {\bibinfo  {journal} {Science}\ }\textbf {\bibinfo {volume} {285}},\ \bibinfo
  {pages} {571} (\bibinfo {year} {1999})}\BibitemShut {NoStop}%
\bibitem [{\citenamefont {Sadler}\ \emph {et~al.}(2006)\citenamefont {Sadler},
  \citenamefont {Higbie}, \citenamefont {Leslie}, \citenamefont
  {Vengalattore},\ and\ \citenamefont {Stamper-Kurn}}]{Sadler2006}%
  \BibitemOpen
  \bibfield  {author} {\bibinfo {author} {\bibfnamefont {L.~E.}\ \bibnamefont
  {Sadler}}, \bibinfo {author} {\bibfnamefont {J.~M.}\ \bibnamefont {Higbie}},
  \bibinfo {author} {\bibfnamefont {S.~R.}\ \bibnamefont {Leslie}}, \bibinfo
  {author} {\bibfnamefont {M.}~\bibnamefont {Vengalattore}},\ and\ \bibinfo
  {author} {\bibfnamefont {D.~M.}\ \bibnamefont {Stamper-Kurn}},\ }\bibfield
  {title} {\bibinfo {title} {Spontaneous symmetry breaking in a quenched
  ferromagnetic spinor {Bose}–{Einstein} condensate},\ }\href
  {https://doi.org/10.1038/nature05094} {\bibfield  {journal} {\bibinfo
  {journal} {Nature}\ }\textbf {\bibinfo {volume} {443}},\ \bibinfo {pages}
  {312} (\bibinfo {year} {2006})}\BibitemShut {NoStop}%
\bibitem [{\citenamefont {Luo}\ \emph {et~al.}(2017)\citenamefont {Luo},
  \citenamefont {Zou}, \citenamefont {Wu}, \citenamefont {Liu}, \citenamefont
  {Han}, \citenamefont {Tey},\ and\ \citenamefont {You}}]{Luo2017}%
  \BibitemOpen
  \bibfield  {author} {\bibinfo {author} {\bibfnamefont {X.-Y.}\ \bibnamefont
  {Luo}}, \bibinfo {author} {\bibfnamefont {Y.-Q.}\ \bibnamefont {Zou}},
  \bibinfo {author} {\bibfnamefont {L.-N.}\ \bibnamefont {Wu}}, \bibinfo
  {author} {\bibfnamefont {Q.}~\bibnamefont {Liu}}, \bibinfo {author}
  {\bibfnamefont {M.-F.}\ \bibnamefont {Han}}, \bibinfo {author} {\bibfnamefont
  {M.~K.}\ \bibnamefont {Tey}},\ and\ \bibinfo {author} {\bibfnamefont
  {L.}~\bibnamefont {You}},\ }\bibfield  {title} {\bibinfo {title}
  {Deterministic entanglement generation from driving through quantum phase
  transitions},\ }\href {https://doi.org/10.1126/science.aag1106} {\bibfield
  {journal} {\bibinfo  {journal} {Science}\ }\textbf {\bibinfo {volume}
  {355}},\ \bibinfo {pages} {620} (\bibinfo {year} {2017})}\BibitemShut
  {NoStop}%
\bibitem [{\citenamefont {Baumann}\ \emph {et~al.}(2010)\citenamefont
  {Baumann}, \citenamefont {Guerlin}, \citenamefont {Brennecke},\ and\
  \citenamefont {Esslinger}}]{Baumann2010}%
  \BibitemOpen
  \bibfield  {author} {\bibinfo {author} {\bibfnamefont {K.}~\bibnamefont
  {Baumann}}, \bibinfo {author} {\bibfnamefont {C.}~\bibnamefont {Guerlin}},
  \bibinfo {author} {\bibfnamefont {F.}~\bibnamefont {Brennecke}},\ and\
  \bibinfo {author} {\bibfnamefont {T.}~\bibnamefont {Esslinger}},\ }\bibfield
  {title} {\bibinfo {title} {Dicke quantum phase transition with a superfluid
  gas in an optical cavity},\ }\href {https://doi.org/10.1038/nature09009}
  {\bibfield  {journal} {\bibinfo  {journal} {Nature}\ }\textbf {\bibinfo
  {volume} {464}},\ \bibinfo {pages} {1301} (\bibinfo {year}
  {2010})}\BibitemShut {NoStop}%
\bibitem [{\citenamefont {Greiner}\ \emph {et~al.}(2002)\citenamefont
  {Greiner}, \citenamefont {Mandel}, \citenamefont {Esslinger}, \citenamefont
  {Hänsch},\ and\ \citenamefont {Bloch}}]{Greiner2002}%
  \BibitemOpen
  \bibfield  {author} {\bibinfo {author} {\bibfnamefont {M.}~\bibnamefont
  {Greiner}}, \bibinfo {author} {\bibfnamefont {O.}~\bibnamefont {Mandel}},
  \bibinfo {author} {\bibfnamefont {T.}~\bibnamefont {Esslinger}}, \bibinfo
  {author} {\bibfnamefont {T.~W.}\ \bibnamefont {Hänsch}},\ and\ \bibinfo
  {author} {\bibfnamefont {I.}~\bibnamefont {Bloch}},\ }\bibfield  {title}
  {\bibinfo {title} {Quantum phase transition from a superfluid to a {Mott}
  insulator in a gas of ultracold atoms},\ }\href
  {https://doi.org/10.1038/415039a} {\bibfield  {journal} {\bibinfo  {journal}
  {Nature}\ }\textbf {\bibinfo {volume} {415}},\ \bibinfo {pages} {39}
  (\bibinfo {year} {2002})}\BibitemShut {NoStop}%
\bibitem [{\citenamefont {Ospelkaus}\ \emph {et~al.}(2010)\citenamefont
  {Ospelkaus}, \citenamefont {Ni}, \citenamefont {Wang}, \citenamefont
  {de~Miranda}, \citenamefont {Neyenhuis}, \citenamefont {Qu{\'e}m{\'e}ner},
  \citenamefont {Julienne}, \citenamefont {Bohn}, \citenamefont {Jin},\ and\
  \citenamefont {Ye}}]{Ospelkaus2010}%
  \BibitemOpen
  \bibfield  {author} {\bibinfo {author} {\bibfnamefont {S.}~\bibnamefont
  {Ospelkaus}}, \bibinfo {author} {\bibfnamefont {K.-K.}\ \bibnamefont {Ni}},
  \bibinfo {author} {\bibfnamefont {D.}~\bibnamefont {Wang}}, \bibinfo {author}
  {\bibfnamefont {M.~H.~G.}\ \bibnamefont {de~Miranda}}, \bibinfo {author}
  {\bibfnamefont {B.}~\bibnamefont {Neyenhuis}}, \bibinfo {author}
  {\bibfnamefont {G.}~\bibnamefont {Qu{\'e}m{\'e}ner}}, \bibinfo {author}
  {\bibfnamefont {P.~S.}\ \bibnamefont {Julienne}}, \bibinfo {author}
  {\bibfnamefont {J.~L.}\ \bibnamefont {Bohn}}, \bibinfo {author}
  {\bibfnamefont {D.~S.}\ \bibnamefont {Jin}},\ and\ \bibinfo {author}
  {\bibfnamefont {J.}~\bibnamefont {Ye}},\ }\bibfield  {title} {\bibinfo
  {title} {Quantum-state controlled chemical reactions of ultracold
  potassium-rubidium molecules},\ }\href
  {https://doi.org/10.1126/science.1184121} {\bibfield  {journal} {\bibinfo
  {journal} {Science}\ }\textbf {\bibinfo {volume} {327}},\ \bibinfo {pages}
  {853} (\bibinfo {year} {2010})}\BibitemShut {NoStop}%
\bibitem [{\citenamefont {Croft}\ \emph {et~al.}(2017)\citenamefont {Croft},
  \citenamefont {Makrides}, \citenamefont {Li}, \citenamefont {Petrov},
  \citenamefont {Kendrick}, \citenamefont {Balakrishnan},\ and\ \citenamefont
  {Kotochigova}}]{Croft2017}%
  \BibitemOpen
  \bibfield  {author} {\bibinfo {author} {\bibfnamefont {J.~F.~E.}\
  \bibnamefont {Croft}}, \bibinfo {author} {\bibfnamefont {C.}~\bibnamefont
  {Makrides}}, \bibinfo {author} {\bibfnamefont {M.}~\bibnamefont {Li}},
  \bibinfo {author} {\bibfnamefont {A.}~\bibnamefont {Petrov}}, \bibinfo
  {author} {\bibfnamefont {B.~K.}\ \bibnamefont {Kendrick}}, \bibinfo {author}
  {\bibfnamefont {N.}~\bibnamefont {Balakrishnan}},\ and\ \bibinfo {author}
  {\bibfnamefont {S.}~\bibnamefont {Kotochigova}},\ }\bibfield  {title}
  {\bibinfo {title} {Universality and chaoticity in ultracold {K+KRb} chemical
  reactions},\ }\href {https://doi.org/10.1038/ncomms15897} {\bibfield
  {journal} {\bibinfo  {journal} {Nat. Commun.}\ }\textbf {\bibinfo {volume}
  {8}},\ \bibinfo {pages} {15897} (\bibinfo {year} {2017})}\BibitemShut
  {NoStop}%
\bibitem [{\citenamefont {Hu}\ \emph {et~al.}(2014)\citenamefont {Hu},
  \citenamefont {Bloom}, \citenamefont {Jin},\ and\ \citenamefont
  {Goldwin}}]{Hu2014}%
  \BibitemOpen
  \bibfield  {author} {\bibinfo {author} {\bibfnamefont {M.-G.}\ \bibnamefont
  {Hu}}, \bibinfo {author} {\bibfnamefont {R.~S.}\ \bibnamefont {Bloom}},
  \bibinfo {author} {\bibfnamefont {D.~S.}\ \bibnamefont {Jin}},\ and\ \bibinfo
  {author} {\bibfnamefont {J.~M.}\ \bibnamefont {Goldwin}},\ }\bibfield
  {title} {\bibinfo {title} {Avalanche-mechanism loss at an atom-molecule
  efimov resonance},\ }\href {https://doi.org/10.1103/PhysRevA.90.013619}
  {\bibfield  {journal} {\bibinfo  {journal} {Phys. Rev. A}\ }\textbf {\bibinfo
  {volume} {90}},\ \bibinfo {pages} {013619} (\bibinfo {year}
  {2014})}\BibitemShut {NoStop}%
\bibitem [{\citenamefont {Ludlow}\ \emph {et~al.}(2015)\citenamefont {Ludlow},
  \citenamefont {Boyd}, \citenamefont {Ye}, \citenamefont {Peik},\ and\
  \citenamefont {Schmidt}}]{Ludlow2015}%
  \BibitemOpen
  \bibfield  {author} {\bibinfo {author} {\bibfnamefont {A.~D.}\ \bibnamefont
  {Ludlow}}, \bibinfo {author} {\bibfnamefont {M.~M.}\ \bibnamefont {Boyd}},
  \bibinfo {author} {\bibfnamefont {J.}~\bibnamefont {Ye}}, \bibinfo {author}
  {\bibfnamefont {E.}~\bibnamefont {Peik}},\ and\ \bibinfo {author}
  {\bibfnamefont {P.~O.}\ \bibnamefont {Schmidt}},\ }\bibfield  {title}
  {\bibinfo {title} {Optical atomic clocks},\ }\href
  {https://doi.org/10.1103/RevModPhys.87.637} {\bibfield  {journal} {\bibinfo
  {journal} {Rev. Mod. Phys.}\ }\textbf {\bibinfo {volume} {87}},\ \bibinfo
  {pages} {637} (\bibinfo {year} {2015})}\BibitemShut {NoStop}%
\bibitem [{\citenamefont {Wu}\ \emph {et~al.}(2019)\citenamefont {Wu},
  \citenamefont {Pagel}, \citenamefont {Malek}, \citenamefont {Nguyen},
  \citenamefont {Zi}, \citenamefont {Scheirer},\ and\ \citenamefont
  {Müller}}]{Wu2019}%
  \BibitemOpen
  \bibfield  {author} {\bibinfo {author} {\bibfnamefont {X.}~\bibnamefont
  {Wu}}, \bibinfo {author} {\bibfnamefont {Z.}~\bibnamefont {Pagel}}, \bibinfo
  {author} {\bibfnamefont {B.~S.}\ \bibnamefont {Malek}}, \bibinfo {author}
  {\bibfnamefont {T.~H.}\ \bibnamefont {Nguyen}}, \bibinfo {author}
  {\bibfnamefont {F.}~\bibnamefont {Zi}}, \bibinfo {author} {\bibfnamefont
  {D.~S.}\ \bibnamefont {Scheirer}},\ and\ \bibinfo {author} {\bibfnamefont
  {H.}~\bibnamefont {Müller}},\ }\bibfield  {title} {\bibinfo {title} {Gravity
  surveys using a mobile atom interferometer},\ }\href
  {https://doi.org/10.1126/sciadv.aax0800} {\bibfield  {journal} {\bibinfo
  {journal} {Sci. Adv.}\ }\textbf {\bibinfo {volume} {5}},\ \bibinfo {pages}
  {eaax0800} (\bibinfo {year} {2019})}\BibitemShut {NoStop}%
\bibitem [{\citenamefont {Hardman}\ \emph {et~al.}(2016)\citenamefont
  {Hardman}, \citenamefont {Everitt}, \citenamefont {McDonald}, \citenamefont
  {Manju}, \citenamefont {Wigley}, \citenamefont {Sooriyabandara},
  \citenamefont {Kuhn}, \citenamefont {Debs}, \citenamefont {Close},\ and\
  \citenamefont {Robins}}]{Hardman2016}%
  \BibitemOpen
  \bibfield  {author} {\bibinfo {author} {\bibfnamefont {K.}~\bibnamefont
  {Hardman}}, \bibinfo {author} {\bibfnamefont {P.}~\bibnamefont {Everitt}},
  \bibinfo {author} {\bibfnamefont {G.}~\bibnamefont {McDonald}}, \bibinfo
  {author} {\bibfnamefont {P.}~\bibnamefont {Manju}}, \bibinfo {author}
  {\bibfnamefont {P.}~\bibnamefont {Wigley}}, \bibinfo {author} {\bibfnamefont
  {M.}~\bibnamefont {Sooriyabandara}}, \bibinfo {author} {\bibfnamefont
  {C.}~\bibnamefont {Kuhn}}, \bibinfo {author} {\bibfnamefont {J.}~\bibnamefont
  {Debs}}, \bibinfo {author} {\bibfnamefont {J.}~\bibnamefont {Close}},\ and\
  \bibinfo {author} {\bibfnamefont {N.}~\bibnamefont {Robins}},\ }\bibfield
  {title} {\bibinfo {title} {Simultaneous {Precision} {Gravimetry} and
  {Magnetic} {Gradiometry} with a {Bose}-{Einstein} {Condensate}: {A} {High}
  {Precision}, {Quantum} {Sensor}},\ }\href
  {https://doi.org/10.1103/PhysRevLett.117.138501} {\bibfield  {journal}
  {\bibinfo  {journal} {Phys. Rev. Lett.}\ }\textbf {\bibinfo {volume} {117}},\
  \bibinfo {pages} {138501} (\bibinfo {year} {2016})},\ \bibinfo {note}
  {publisher: American Physical Society}\BibitemShut {NoStop}%
\bibitem [{\citenamefont {Freier}\ \emph {et~al.}(2016)\citenamefont {Freier},
  \citenamefont {Hauth}, \citenamefont {Schkolnik}, \citenamefont {Leykauf},
  \citenamefont {Schilling}, \citenamefont {Wziontek}, \citenamefont
  {Scherneck}, \citenamefont {Müller},\ and\ \citenamefont
  {Peters}}]{Freier2016}%
  \BibitemOpen
  \bibfield  {author} {\bibinfo {author} {\bibfnamefont {C.}~\bibnamefont
  {Freier}}, \bibinfo {author} {\bibfnamefont {M.}~\bibnamefont {Hauth}},
  \bibinfo {author} {\bibfnamefont {V.}~\bibnamefont {Schkolnik}}, \bibinfo
  {author} {\bibfnamefont {B.}~\bibnamefont {Leykauf}}, \bibinfo {author}
  {\bibfnamefont {M.}~\bibnamefont {Schilling}}, \bibinfo {author}
  {\bibfnamefont {H.}~\bibnamefont {Wziontek}}, \bibinfo {author}
  {\bibfnamefont {H.-G.}\ \bibnamefont {Scherneck}}, \bibinfo {author}
  {\bibfnamefont {J.}~\bibnamefont {Müller}},\ and\ \bibinfo {author}
  {\bibfnamefont {A.}~\bibnamefont {Peters}},\ }\bibfield  {title} {\bibinfo
  {title} {Mobile quantum gravity sensor with unprecedented stability},\ }\href
  {https://doi.org/10.1088/1742-6596/723/1/012050} {\bibfield  {journal}
  {\bibinfo  {journal} {J. Phys. Conf. Ser.}\ }\textbf {\bibinfo {volume}
  {723}},\ \bibinfo {pages} {012050} (\bibinfo {year} {2016})}\BibitemShut
  {NoStop}%
\bibitem [{\citenamefont {Ménoret}\ \emph {et~al.}(2018)\citenamefont
  {Ménoret}, \citenamefont {Vermeulen}, \citenamefont {Le~Moigne},
  \citenamefont {Bonvalot}, \citenamefont {Bouyer}, \citenamefont {Landragin},\
  and\ \citenamefont {Desruelle}}]{Menoret2018}%
  \BibitemOpen
  \bibfield  {author} {\bibinfo {author} {\bibfnamefont {V.}~\bibnamefont
  {Ménoret}}, \bibinfo {author} {\bibfnamefont {P.}~\bibnamefont {Vermeulen}},
  \bibinfo {author} {\bibfnamefont {N.}~\bibnamefont {Le~Moigne}}, \bibinfo
  {author} {\bibfnamefont {S.}~\bibnamefont {Bonvalot}}, \bibinfo {author}
  {\bibfnamefont {P.}~\bibnamefont {Bouyer}}, \bibinfo {author} {\bibfnamefont
  {A.}~\bibnamefont {Landragin}},\ and\ \bibinfo {author} {\bibfnamefont
  {B.}~\bibnamefont {Desruelle}},\ }\bibfield  {title} {\bibinfo {title}
  {Gravity measurements below {$10 ^{-9}$} g with a transportable absolute
  quantum gravimeter},\ }\href {https://doi.org/10.1038/s41598-018-30608-1}
  {\bibfield  {journal} {\bibinfo  {journal} {Sci. Rep.}\ }\textbf {\bibinfo
  {volume} {8}},\ \bibinfo {pages} {12300} (\bibinfo {year}
  {2018})}\BibitemShut {NoStop}%
\bibitem [{\citenamefont {Pezz{\`{e}}}\ \emph {et~al.}(2018)\citenamefont
  {Pezz{\`{e}}}, \citenamefont {Smerzi}, \citenamefont {Oberthaler},
  \citenamefont {Schmied},\ and\ \citenamefont {Treutlein}}]{Pezze2018}%
  \BibitemOpen
  \bibfield  {author} {\bibinfo {author} {\bibfnamefont {L.}~\bibnamefont
  {Pezz{\`{e}}}}, \bibinfo {author} {\bibfnamefont {A.}~\bibnamefont {Smerzi}},
  \bibinfo {author} {\bibfnamefont {M.~K.}\ \bibnamefont {Oberthaler}},
  \bibinfo {author} {\bibfnamefont {R.}~\bibnamefont {Schmied}},\ and\ \bibinfo
  {author} {\bibfnamefont {P.}~\bibnamefont {Treutlein}},\ }\bibfield  {title}
  {\bibinfo {title} {Quantum metrology with nonclassical states of atomic
  ensembles},\ }\href {https://doi.org/10.1103/revmodphys.90.035005} {\bibfield
   {journal} {\bibinfo  {journal} {Rev. Mod. Phys.}\ }\textbf {\bibinfo
  {volume} {90}},\ \bibinfo {pages} {035005} (\bibinfo {year}
  {2018})}\BibitemShut {NoStop}%
\bibitem [{\citenamefont {Cox}\ \emph {et~al.}(2016)\citenamefont {Cox},
  \citenamefont {Greve}, \citenamefont {Weiner},\ and\ \citenamefont
  {Thompson}}]{Cox2016}%
  \BibitemOpen
  \bibfield  {author} {\bibinfo {author} {\bibfnamefont {K.~C.}\ \bibnamefont
  {Cox}}, \bibinfo {author} {\bibfnamefont {G.~P.}\ \bibnamefont {Greve}},
  \bibinfo {author} {\bibfnamefont {J.~M.}\ \bibnamefont {Weiner}},\ and\
  \bibinfo {author} {\bibfnamefont {J.~K.}\ \bibnamefont {Thompson}},\
  }\bibfield  {title} {\bibinfo {title} {Deterministic {Squeezed} {States} with
  {Collective} {Measurements} and {Feedback}},\ }\href
  {https://doi.org/10.1103/PhysRevLett.116.093602} {\bibfield  {journal}
  {\bibinfo  {journal} {Phys. Rev. Lett.}\ }\textbf {\bibinfo {volume} {116}},\
  \bibinfo {pages} {093602} (\bibinfo {year} {2016})}\BibitemShut {NoStop}%
\bibitem [{\citenamefont {Kohlhaas}\ \emph {et~al.}(2015)\citenamefont
  {Kohlhaas}, \citenamefont {Bertoldi}, \citenamefont {Cantin}, \citenamefont
  {Aspect}, \citenamefont {Landragin},\ and\ \citenamefont
  {Bouyer}}]{Kohlhaas2015}%
  \BibitemOpen
  \bibfield  {author} {\bibinfo {author} {\bibfnamefont {R.}~\bibnamefont
  {Kohlhaas}}, \bibinfo {author} {\bibfnamefont {A.}~\bibnamefont {Bertoldi}},
  \bibinfo {author} {\bibfnamefont {E.}~\bibnamefont {Cantin}}, \bibinfo
  {author} {\bibfnamefont {A.}~\bibnamefont {Aspect}}, \bibinfo {author}
  {\bibfnamefont {A.}~\bibnamefont {Landragin}},\ and\ \bibinfo {author}
  {\bibfnamefont {P.}~\bibnamefont {Bouyer}},\ }\bibfield  {title} {\bibinfo
  {title} {Phase {Locking} a {Clock} {Oscillator} to a {Coherent} {Atomic}
  {Ensemble}},\ }\href {https://doi.org/10.1103/PhysRevX.5.021011} {\bibfield
  {journal} {\bibinfo  {journal} {Phys. Rev. X}\ }\textbf {\bibinfo {volume}
  {5}},\ \bibinfo {pages} {021011} (\bibinfo {year} {2015})}\BibitemShut
  {NoStop}%
\bibitem [{\citenamefont {Horvath}\ \emph {et~al.}(2017)\citenamefont
  {Horvath}, \citenamefont {Thomas}, \citenamefont {Tiesinga}, \citenamefont
  {Deb},\ and\ \citenamefont {Kj{\ae}rgaard}}]{Horvath2017}%
  \BibitemOpen
  \bibfield  {author} {\bibinfo {author} {\bibfnamefont {M.~S.~J.}\
  \bibnamefont {Horvath}}, \bibinfo {author} {\bibfnamefont {R.}~\bibnamefont
  {Thomas}}, \bibinfo {author} {\bibfnamefont {E.}~\bibnamefont {Tiesinga}},
  \bibinfo {author} {\bibfnamefont {A.~B.}\ \bibnamefont {Deb}},\ and\ \bibinfo
  {author} {\bibfnamefont {N.}~\bibnamefont {Kj{\ae}rgaard}},\ }\bibfield
  {title} {\bibinfo {title} {Above-threshold scattering about a {Feshbach}
  resonance for ultracold atoms in an optical collider},\ }\href
  {https://doi.org/10.1038/s41467-017-00458-y} {\bibfield  {journal} {\bibinfo
  {journal} {Nat. Commun.}\ }\textbf {\bibinfo {volume} {8}},\ \bibinfo {pages}
  {452} (\bibinfo {year} {2017})}\BibitemShut {NoStop}%
\bibitem [{\citenamefont {Thomas}\ \emph {et~al.}(2016)\citenamefont {Thomas},
  \citenamefont {Roberts}, \citenamefont {Tiesinga}, \citenamefont {Wade},
  \citenamefont {Blakie}, \citenamefont {Deb},\ and\ \citenamefont
  {Kj{\ae}rgaard}}]{Thomas2016}%
  \BibitemOpen
  \bibfield  {author} {\bibinfo {author} {\bibfnamefont {R.}~\bibnamefont
  {Thomas}}, \bibinfo {author} {\bibfnamefont {K.~O.}\ \bibnamefont {Roberts}},
  \bibinfo {author} {\bibfnamefont {E.}~\bibnamefont {Tiesinga}}, \bibinfo
  {author} {\bibfnamefont {A.~C.~J.}\ \bibnamefont {Wade}}, \bibinfo {author}
  {\bibfnamefont {P.~B.}\ \bibnamefont {Blakie}}, \bibinfo {author}
  {\bibfnamefont {A.~B.}\ \bibnamefont {Deb}},\ and\ \bibinfo {author}
  {\bibfnamefont {N.}~\bibnamefont {Kj{\ae}rgaard}},\ }\bibfield  {title}
  {\bibinfo {title} {Multiple scattering dynamics of fermions at an isolated
  p-wave resonance},\ }\href {https://doi.org/10.1038/ncomms12069} {\bibfield
  {journal} {\bibinfo  {journal} {Nat. Commun.}\ }\textbf {\bibinfo {volume}
  {7}},\ \bibinfo {pages} {12069} (\bibinfo {year} {2016})}\BibitemShut
  {NoStop}%
\bibitem [{\citenamefont {Thomas}\ \emph {et~al.}(2018)\citenamefont {Thomas},
  \citenamefont {Chilcott}, \citenamefont {Tiesinga}, \citenamefont {Deb},\
  and\ \citenamefont {Kj{\ae}rgaard}}]{Thomas2018}%
  \BibitemOpen
  \bibfield  {author} {\bibinfo {author} {\bibfnamefont {R.}~\bibnamefont
  {Thomas}}, \bibinfo {author} {\bibfnamefont {M.}~\bibnamefont {Chilcott}},
  \bibinfo {author} {\bibfnamefont {E.}~\bibnamefont {Tiesinga}}, \bibinfo
  {author} {\bibfnamefont {A.~B.}\ \bibnamefont {Deb}},\ and\ \bibinfo {author}
  {\bibfnamefont {N.}~\bibnamefont {Kj{\ae}rgaard}},\ }\bibfield  {title}
  {\bibinfo {title} {Observation of bound state self-interaction in a nano-{eV}
  atom collider},\ }\href {https://doi.org/10.1038/s41467-018-07375-8}
  {\bibfield  {journal} {\bibinfo  {journal} {Nat. Commun.}\ }\textbf {\bibinfo
  {volume} {9}},\ \bibinfo {pages} {4895} (\bibinfo {year} {2018})}\BibitemShut
  {NoStop}%
\bibitem [{\citenamefont {Koch}\ \emph {et~al.}(2008)\citenamefont {Koch},
  \citenamefont {Lahaye}, \citenamefont {Metz}, \citenamefont {Fröhlich},
  \citenamefont {Griesmaier},\ and\ \citenamefont {Pfau}}]{Koch2008}%
  \BibitemOpen
  \bibfield  {author} {\bibinfo {author} {\bibfnamefont {T.}~\bibnamefont
  {Koch}}, \bibinfo {author} {\bibfnamefont {T.}~\bibnamefont {Lahaye}},
  \bibinfo {author} {\bibfnamefont {J.}~\bibnamefont {Metz}}, \bibinfo {author}
  {\bibfnamefont {B.}~\bibnamefont {Fröhlich}}, \bibinfo {author}
  {\bibfnamefont {A.}~\bibnamefont {Griesmaier}},\ and\ \bibinfo {author}
  {\bibfnamefont {T.}~\bibnamefont {Pfau}},\ }\bibfield  {title} {\bibinfo
  {title} {Stabilization of a purely dipolar quantum gas against collapse},\
  }\href {https://doi.org/10.1038/nphys887} {\bibfield  {journal} {\bibinfo
  {journal} {Nat. Phys.}\ }\textbf {\bibinfo {volume} {4}},\ \bibinfo {pages}
  {218} (\bibinfo {year} {2008})}\BibitemShut {NoStop}%
\bibitem [{\citenamefont {Chomaz}\ \emph {et~al.}(2019)\citenamefont {Chomaz},
  \citenamefont {Petter}, \citenamefont {Ilzhöfer}, \citenamefont {Natale},
  \citenamefont {Trautmann}, \citenamefont {Politi}, \citenamefont
  {Durastante}, \citenamefont {van Bijnen}, \citenamefont {Patscheider},
  \citenamefont {Sohmen}, \citenamefont {Mark},\ and\ \citenamefont
  {Ferlaino}}]{Chomaz2019}%
  \BibitemOpen
  \bibfield  {author} {\bibinfo {author} {\bibfnamefont {L.}~\bibnamefont
  {Chomaz}}, \bibinfo {author} {\bibfnamefont {D.}~\bibnamefont {Petter}},
  \bibinfo {author} {\bibfnamefont {P.}~\bibnamefont {Ilzhöfer}}, \bibinfo
  {author} {\bibfnamefont {G.}~\bibnamefont {Natale}}, \bibinfo {author}
  {\bibfnamefont {A.}~\bibnamefont {Trautmann}}, \bibinfo {author}
  {\bibfnamefont {C.}~\bibnamefont {Politi}}, \bibinfo {author} {\bibfnamefont
  {G.}~\bibnamefont {Durastante}}, \bibinfo {author} {\bibfnamefont
  {R.}~\bibnamefont {van Bijnen}}, \bibinfo {author} {\bibfnamefont
  {A.}~\bibnamefont {Patscheider}}, \bibinfo {author} {\bibfnamefont
  {M.}~\bibnamefont {Sohmen}}, \bibinfo {author} {\bibfnamefont
  {M.}~\bibnamefont {Mark}},\ and\ \bibinfo {author} {\bibfnamefont
  {F.}~\bibnamefont {Ferlaino}},\ }\bibfield  {title} {\bibinfo {title}
  {Long-{Lived} and {Transient} {Supersolid} {Behaviors} in {Dipolar} {Quantum}
  {Gases}},\ }\href {https://doi.org/10.1103/PhysRevX.9.021012} {\bibfield
  {journal} {\bibinfo  {journal} {Phys. Rev. X}\ }\textbf {\bibinfo {volume}
  {9}},\ \bibinfo {pages} {021012} (\bibinfo {year} {2019})},\ \bibinfo {note}
  {publisher: American Physical Society}\BibitemShut {NoStop}%
\bibitem [{\citenamefont {Lewandowski}\ \emph {et~al.}(2003)\citenamefont
  {Lewandowski}, \citenamefont {Harber}, \citenamefont {Whitaker},\ and\
  \citenamefont {Cornell}}]{Lewandowski2003}%
  \BibitemOpen
  \bibfield  {author} {\bibinfo {author} {\bibfnamefont {H.~J.}\ \bibnamefont
  {Lewandowski}}, \bibinfo {author} {\bibfnamefont {D.~M.}\ \bibnamefont
  {Harber}}, \bibinfo {author} {\bibfnamefont {D.~L.}\ \bibnamefont
  {Whitaker}},\ and\ \bibinfo {author} {\bibfnamefont {E.~A.}\ \bibnamefont
  {Cornell}},\ }\bibfield  {title} {\bibinfo {title} {Simplified {System} for
  {Creating} a {Bose}–{Einstein} {Condensate}},\ }\href
  {https://doi.org/10.1023/A:1024800600621} {\bibfield  {journal} {\bibinfo
  {journal} {J. Low Temp. Phys.}\ }\textbf {\bibinfo {volume} {132}},\ \bibinfo
  {pages} {309} (\bibinfo {year} {2003})}\BibitemShut {NoStop}%
\bibitem [{\citenamefont {Streed}\ \emph {et~al.}(2006)\citenamefont {Streed},
  \citenamefont {Chikkatur}, \citenamefont {Gustavson}, \citenamefont {Boyd},
  \citenamefont {Torii}, \citenamefont {Schneble}, \citenamefont {Campbell},
  \citenamefont {Pritchard},\ and\ \citenamefont {Ketterle}}]{Streed2006}%
  \BibitemOpen
  \bibfield  {author} {\bibinfo {author} {\bibfnamefont {E.~W.}\ \bibnamefont
  {Streed}}, \bibinfo {author} {\bibfnamefont {A.~P.}\ \bibnamefont
  {Chikkatur}}, \bibinfo {author} {\bibfnamefont {T.~L.}\ \bibnamefont
  {Gustavson}}, \bibinfo {author} {\bibfnamefont {M.}~\bibnamefont {Boyd}},
  \bibinfo {author} {\bibfnamefont {Y.}~\bibnamefont {Torii}}, \bibinfo
  {author} {\bibfnamefont {D.}~\bibnamefont {Schneble}}, \bibinfo {author}
  {\bibfnamefont {G.~K.}\ \bibnamefont {Campbell}}, \bibinfo {author}
  {\bibfnamefont {D.~E.}\ \bibnamefont {Pritchard}},\ and\ \bibinfo {author}
  {\bibfnamefont {W.}~\bibnamefont {Ketterle}},\ }\bibfield  {title} {\bibinfo
  {title} {Large atom number {Bose}-{Einstein} condensate machines},\ }\href
  {https://doi.org/10.1063/1.2163977} {\bibfield  {journal} {\bibinfo
  {journal} {Rev. Sci. Instrum.}\ }\textbf {\bibinfo {volume} {77}},\ \bibinfo
  {pages} {023106} (\bibinfo {year} {2006})}\BibitemShut {NoStop}%
\bibitem [{\citenamefont {Chilcott}\ and\ \citenamefont
  {Kj{\ae}rgaard}(2021)}]{Chilcott2021}%
  \BibitemOpen
  \bibfield  {author} {\bibinfo {author} {\bibfnamefont {M.}~\bibnamefont
  {Chilcott}}\ and\ \bibinfo {author} {\bibfnamefont {N.}~\bibnamefont
  {Kj{\ae}rgaard}},\ }\bibfield  {title} {\bibinfo {title} {Low-cost wireless
  condition monitoring system for an ultracold atom machine},\ }\href
  {https://doi.org/10.1016/j.iot.2020.100345} {\bibfield  {journal} {\bibinfo
  {journal} {Internet of Things}\ }\textbf {\bibinfo {volume} {13}},\ \bibinfo
  {pages} {100345} (\bibinfo {year} {2021})}\BibitemShut {NoStop}%
\bibitem [{\citenamefont {Sawyer}\ \emph {et~al.}(2012)\citenamefont {Sawyer},
  \citenamefont {Deb}, \citenamefont {McKellar},\ and\ \citenamefont
  {Kj\ae{}rgaard}}]{Sawyer2012}%
  \BibitemOpen
  \bibfield  {author} {\bibinfo {author} {\bibfnamefont {B.~J.}\ \bibnamefont
  {Sawyer}}, \bibinfo {author} {\bibfnamefont {A.~B.}\ \bibnamefont {Deb}},
  \bibinfo {author} {\bibfnamefont {T.}~\bibnamefont {McKellar}},\ and\
  \bibinfo {author} {\bibfnamefont {N.}~\bibnamefont {Kj\ae{}rgaard}},\
  }\bibfield  {title} {\bibinfo {title} {Reducing number fluctuations of
  ultracold atomic gases via dispersive interrogation},\ }\href
  {https://doi.org/10.1103/PhysRevA.86.065401} {\bibfield  {journal} {\bibinfo
  {journal} {Phys. Rev. A}\ }\textbf {\bibinfo {volume} {86}},\ \bibinfo
  {pages} {065401} (\bibinfo {year} {2012})}\BibitemShut {NoStop}%
\bibitem [{\citenamefont {Lye}\ \emph {et~al.}(2003)\citenamefont {Lye},
  \citenamefont {Hope},\ and\ \citenamefont {Close}}]{Lye2003}%
  \BibitemOpen
  \bibfield  {author} {\bibinfo {author} {\bibfnamefont {J.~E.}\ \bibnamefont
  {Lye}}, \bibinfo {author} {\bibfnamefont {J.~J.}\ \bibnamefont {Hope}},\ and\
  \bibinfo {author} {\bibfnamefont {J.~D.}\ \bibnamefont {Close}},\ }\bibfield
  {title} {\bibinfo {title} {Nondestructive dynamic detectors for
  {Bose}-{Einstein} condensates},\ }\href
  {https://doi.org/10.1103/PhysRevA.67.043609} {\bibfield  {journal} {\bibinfo
  {journal} {Phys. Rev. A}\ }\textbf {\bibinfo {volume} {67}},\ \bibinfo
  {pages} {043609} (\bibinfo {year} {2003})}\BibitemShut {NoStop}%
\bibitem [{\citenamefont {Gajdacz}\ \emph {et~al.}(2016)\citenamefont
  {Gajdacz}, \citenamefont {Hilliard}, \citenamefont {Kristensen},
  \citenamefont {Pedersen}, \citenamefont {Klempt}, \citenamefont {Arlt},\ and\
  \citenamefont {Sherson}}]{Gajdacz2016}%
  \BibitemOpen
  \bibfield  {author} {\bibinfo {author} {\bibfnamefont {M.}~\bibnamefont
  {Gajdacz}}, \bibinfo {author} {\bibfnamefont {A.~J.}\ \bibnamefont
  {Hilliard}}, \bibinfo {author} {\bibfnamefont {M.~A.}\ \bibnamefont
  {Kristensen}}, \bibinfo {author} {\bibfnamefont {P.~L.}\ \bibnamefont
  {Pedersen}}, \bibinfo {author} {\bibfnamefont {C.}~\bibnamefont {Klempt}},
  \bibinfo {author} {\bibfnamefont {J.~J.}\ \bibnamefont {Arlt}},\ and\
  \bibinfo {author} {\bibfnamefont {J.~F.}\ \bibnamefont {Sherson}},\
  }\bibfield  {title} {\bibinfo {title} {Preparation of ultracold atom clouds
  at the shot noise level},\ }\href
  {https://doi.org/10.1103/PhysRevLett.117.073604} {\bibfield  {journal}
  {\bibinfo  {journal} {Phys. Rev. Lett.}\ }\textbf {\bibinfo {volume} {117}},\
  \bibinfo {pages} {073604} (\bibinfo {year} {2016})}\BibitemShut {NoStop}%
\bibitem [{\citenamefont {Kristensen}\ \emph {et~al.}(2019)\citenamefont
  {Kristensen}, \citenamefont {Christensen}, \citenamefont {Gajdacz},
  \citenamefont {Iglicki}, \citenamefont {Paw\l{}owski}, \citenamefont
  {Klempt}, \citenamefont {Sherson}, \citenamefont {Rzazewski}, \citenamefont
  {Hilliard},\ and\ \citenamefont {Arlt}}]{Kristensen2019}%
  \BibitemOpen
  \bibfield  {author} {\bibinfo {author} {\bibfnamefont {M.~A.}\ \bibnamefont
  {Kristensen}}, \bibinfo {author} {\bibfnamefont {M.~B.}\ \bibnamefont
  {Christensen}}, \bibinfo {author} {\bibfnamefont {M.}~\bibnamefont
  {Gajdacz}}, \bibinfo {author} {\bibfnamefont {M.}~\bibnamefont {Iglicki}},
  \bibinfo {author} {\bibfnamefont {K.}~\bibnamefont {Paw\l{}owski}}, \bibinfo
  {author} {\bibfnamefont {C.}~\bibnamefont {Klempt}}, \bibinfo {author}
  {\bibfnamefont {J.~F.}\ \bibnamefont {Sherson}}, \bibinfo {author}
  {\bibfnamefont {K.}~\bibnamefont {Rzazewski}}, \bibinfo {author}
  {\bibfnamefont {A.~J.}\ \bibnamefont {Hilliard}},\ and\ \bibinfo {author}
  {\bibfnamefont {J.~J.}\ \bibnamefont {Arlt}},\ }\bibfield  {title} {\bibinfo
  {title} {Observation of atom number fluctuations in a {Bose-Einstein}
  condensate},\ }\href {https://doi.org/10.1103/PhysRevLett.122.163601}
  {\bibfield  {journal} {\bibinfo  {journal} {Phys. Rev. Lett.}\ }\textbf
  {\bibinfo {volume} {122}},\ \bibinfo {pages} {163601} (\bibinfo {year}
  {2019})}\BibitemShut {NoStop}%
\bibitem [{\citenamefont {Christensen}\ \emph {et~al.}()\citenamefont
  {Christensen}, \citenamefont {Vibel}, \citenamefont {Hilliard}, \citenamefont
  {Kruk}, \citenamefont {Pawłowski}, \citenamefont {Hryniuk}, \citenamefont
  {Rzążewski}, \citenamefont {Kristensen},\ and\ \citenamefont
  {Arlt}}]{Christensen2020}%
  \BibitemOpen
  \bibfield  {author} {\bibinfo {author} {\bibfnamefont {M.~B.}\ \bibnamefont
  {Christensen}}, \bibinfo {author} {\bibfnamefont {T.}~\bibnamefont {Vibel}},
  \bibinfo {author} {\bibfnamefont {A.~J.}\ \bibnamefont {Hilliard}}, \bibinfo
  {author} {\bibfnamefont {M.~B.}\ \bibnamefont {Kruk}}, \bibinfo {author}
  {\bibfnamefont {K.}~\bibnamefont {Pawłowski}}, \bibinfo {author}
  {\bibfnamefont {D.}~\bibnamefont {Hryniuk}}, \bibinfo {author} {\bibfnamefont
  {K.}~\bibnamefont {Rzążewski}}, \bibinfo {author} {\bibfnamefont {M.~A.}\
  \bibnamefont {Kristensen}},\ and\ \bibinfo {author} {\bibfnamefont {J.~J.}\
  \bibnamefont {Arlt}},\ }\bibfield  {title} {\bibinfo {title} {Observation of
  microcanonical atom number fluctuations in a bose-einstein condensate},\
  }\href@noop {} {\ }\Eprint {https://arxiv.org/abs/2011.08736} {2011.08736}
  \BibitemShut {NoStop}%
\bibitem [{\citenamefont {Gajdacz}\ \emph {et~al.}(2013)\citenamefont
  {Gajdacz}, \citenamefont {Pedersen}, \citenamefont {Mørch}, \citenamefont
  {Hilliard}, \citenamefont {Arlt},\ and\ \citenamefont
  {Sherson}}]{Gajdacz2013}%
  \BibitemOpen
  \bibfield  {author} {\bibinfo {author} {\bibfnamefont {M.}~\bibnamefont
  {Gajdacz}}, \bibinfo {author} {\bibfnamefont {P.~L.}\ \bibnamefont
  {Pedersen}}, \bibinfo {author} {\bibfnamefont {T.}~\bibnamefont {Mørch}},
  \bibinfo {author} {\bibfnamefont {A.~J.}\ \bibnamefont {Hilliard}}, \bibinfo
  {author} {\bibfnamefont {J.}~\bibnamefont {Arlt}},\ and\ \bibinfo {author}
  {\bibfnamefont {J.~F.}\ \bibnamefont {Sherson}},\ }\bibfield  {title}
  {\bibinfo {title} {Non-destructive {Faraday} imaging of dynamically
  controlled ultracold atoms},\ }\href {https://doi.org/10.1063/1.4818913}
  {\bibfield  {journal} {\bibinfo  {journal} {Rev. Sci. Instrum.}\ }\textbf
  {\bibinfo {volume} {84}},\ \bibinfo {pages} {083105} (\bibinfo {year}
  {2013})}\BibitemShut {NoStop}%
\bibitem [{\citenamefont {Aveline}\ \emph {et~al.}(2020)\citenamefont
  {Aveline}, \citenamefont {Williams}, \citenamefont {Elliott}, \citenamefont
  {Dutenhoffer}, \citenamefont {Kellogg}, \citenamefont {Kohel}, \citenamefont
  {Lay}, \citenamefont {Oudrhiri}, \citenamefont {Shotwell}, \citenamefont
  {Yu},\ and\ \citenamefont {Thompson}}]{Aveline2020}%
  \BibitemOpen
  \bibfield  {author} {\bibinfo {author} {\bibfnamefont {D.~C.}\ \bibnamefont
  {Aveline}}, \bibinfo {author} {\bibfnamefont {J.~R.}\ \bibnamefont
  {Williams}}, \bibinfo {author} {\bibfnamefont {E.~R.}\ \bibnamefont
  {Elliott}}, \bibinfo {author} {\bibfnamefont {C.}~\bibnamefont
  {Dutenhoffer}}, \bibinfo {author} {\bibfnamefont {J.~R.}\ \bibnamefont
  {Kellogg}}, \bibinfo {author} {\bibfnamefont {J.~M.}\ \bibnamefont {Kohel}},
  \bibinfo {author} {\bibfnamefont {N.~E.}\ \bibnamefont {Lay}}, \bibinfo
  {author} {\bibfnamefont {K.}~\bibnamefont {Oudrhiri}}, \bibinfo {author}
  {\bibfnamefont {R.~F.}\ \bibnamefont {Shotwell}}, \bibinfo {author}
  {\bibfnamefont {N.}~\bibnamefont {Yu}},\ and\ \bibinfo {author}
  {\bibfnamefont {R.~J.}\ \bibnamefont {Thompson}},\ }\bibfield  {title}
  {\bibinfo {title} {Observation of {Bose}–{Einstein} condensates in an
  {Earth}-orbiting research lab},\ }\href
  {https://doi.org/10.1038/s41586-020-2346-1} {\bibfield  {journal} {\bibinfo
  {journal} {Nature}\ }\textbf {\bibinfo {volume} {582}},\ \bibinfo {pages}
  {193} (\bibinfo {year} {2020})}\BibitemShut {NoStop}%
\bibitem [{\citenamefont {Julsgaard}\ \emph {et~al.}(2001)\citenamefont
  {Julsgaard}, \citenamefont {Kozhekin},\ and\ \citenamefont
  {Polzik}}]{Julsgaard2001}%
  \BibitemOpen
  \bibfield  {author} {\bibinfo {author} {\bibfnamefont {B.}~\bibnamefont
  {Julsgaard}}, \bibinfo {author} {\bibfnamefont {A.}~\bibnamefont
  {Kozhekin}},\ and\ \bibinfo {author} {\bibfnamefont {E.~S.}\ \bibnamefont
  {Polzik}},\ }\bibfield  {title} {\bibinfo {title} {Experimental long-lived
  entanglement of two macroscopic objects},\ }\href
  {https://doi.org/10.1038/35096524} {\bibfield  {journal} {\bibinfo  {journal}
  {Nature}\ }\textbf {\bibinfo {volume} {413}},\ \bibinfo {pages} {400}
  (\bibinfo {year} {2001})}\BibitemShut {NoStop}%
\bibitem [{\citenamefont {Kaminski}\ \emph {et~al.}(2012)\citenamefont
  {Kaminski}, \citenamefont {Kampel}, \citenamefont {Steenstrup}, \citenamefont
  {Griesmaier}, \citenamefont {Polzik},\ and\ \citenamefont
  {Müller}}]{Kaminski2012}%
  \BibitemOpen
  \bibfield  {author} {\bibinfo {author} {\bibfnamefont {F.}~\bibnamefont
  {Kaminski}}, \bibinfo {author} {\bibfnamefont {N.~S.}\ \bibnamefont
  {Kampel}}, \bibinfo {author} {\bibfnamefont {M.~P.~H.}\ \bibnamefont
  {Steenstrup}}, \bibinfo {author} {\bibfnamefont {A.}~\bibnamefont
  {Griesmaier}}, \bibinfo {author} {\bibfnamefont {E.~S.}\ \bibnamefont
  {Polzik}},\ and\ \bibinfo {author} {\bibfnamefont {J.~H.}\ \bibnamefont
  {Müller}},\ }\bibfield  {title} {\bibinfo {title} {In-situ dual-port
  polarization contrast imaging of {Faraday} rotation in a high optical depth
  ultracold {$\rm ^{87}Rb$} atomic ensemble},\ }\href
  {https://doi.org/10.1140/epjd/e2012-30038-0} {\bibfield  {journal} {\bibinfo
  {journal} {Eur. Phys. J. D}\ }\textbf {\bibinfo {volume} {66}},\ \bibinfo
  {pages} {227} (\bibinfo {year} {2012})}\BibitemShut {NoStop}%
\bibitem [{\citenamefont {Deb}\ \emph {et~al.}(2020)\citenamefont {Deb},
  \citenamefont {Chung},\ and\ \citenamefont {Kj{\ae}rgaard}}]{Deb2020}%
  \BibitemOpen
  \bibfield  {author} {\bibinfo {author} {\bibfnamefont {A.~B.}\ \bibnamefont
  {Deb}}, \bibinfo {author} {\bibfnamefont {J.}~\bibnamefont {Chung}},\ and\
  \bibinfo {author} {\bibfnamefont {N.}~\bibnamefont {Kj{\ae}rgaard}},\
  }\bibfield  {title} {\bibinfo {title} {Dispersive detection of atomic
  ensembles in the presence of strong lensing},\ }\href
  {https://doi.org/10.1088/1367-2630/ab9553} {\bibfield  {journal} {\bibinfo
  {journal} {New J. Phys.}\ }\textbf {\bibinfo {volume} {22}},\ \bibinfo
  {pages} {073017} (\bibinfo {year} {2020})}\BibitemShut {NoStop}%
\bibitem [{\citenamefont {Hume}\ \emph {et~al.}(2013)\citenamefont {Hume},
  \citenamefont {Stroescu}, \citenamefont {Joos}, \citenamefont {Muessel},
  \citenamefont {Strobel},\ and\ \citenamefont {Oberthaler}}]{Hume2013}%
  \BibitemOpen
  \bibfield  {author} {\bibinfo {author} {\bibfnamefont {D.~B.}\ \bibnamefont
  {Hume}}, \bibinfo {author} {\bibfnamefont {I.}~\bibnamefont {Stroescu}},
  \bibinfo {author} {\bibfnamefont {M.}~\bibnamefont {Joos}}, \bibinfo {author}
  {\bibfnamefont {W.}~\bibnamefont {Muessel}}, \bibinfo {author} {\bibfnamefont
  {H.}~\bibnamefont {Strobel}},\ and\ \bibinfo {author} {\bibfnamefont {M.~K.}\
  \bibnamefont {Oberthaler}},\ }\bibfield  {title} {\bibinfo {title} {Accurate
  {Atom} {Counting} in {Mesoscopic} {Ensembles}},\ }\href
  {https://doi.org/10.1103/PhysRevLett.111.253001} {\bibfield  {journal}
  {\bibinfo  {journal} {Phys. Rev. Lett.}\ }\textbf {\bibinfo {volume} {111}},\
  \bibinfo {pages} {253001} (\bibinfo {year} {2013})}\BibitemShut {NoStop}%
\bibitem [{\citenamefont {Bechhoefer}(2005)}]{Bechhoefer2005}%
  \BibitemOpen
  \bibfield  {author} {\bibinfo {author} {\bibfnamefont {J.}~\bibnamefont
  {Bechhoefer}},\ }\bibfield  {title} {\bibinfo {title} {Feedback for
  physicists: A tutorial essay on control},\ }\href
  {https://doi.org/10.1103/RevModPhys.77.783} {\bibfield  {journal} {\bibinfo
  {journal} {Rev. Mod. Phys.}\ }\textbf {\bibinfo {volume} {77}},\ \bibinfo
  {pages} {783} (\bibinfo {year} {2005})}\BibitemShut {NoStop}%
\bibitem [{IEE(2009)}]{IEEEStandard}%
  \BibitemOpen
  \href {https://doi.org/10.1109/IEEESTD.1999.90575} {\emph {\bibinfo {title}
  {IEEE Standard Definitions of Physical Quantities for Fundamental Frequency
  and Time Metrology-Random Instabilities}}} (\bibinfo {year}
  {2009})\BibitemShut {NoStop}%
\bibitem [{Exc()}]{ExcessNoise}%
  \BibitemOpen
  \href@noop {} {}\bibinfo {note} {APDs experience excess noise above the
  photon shot noise which is expressed as a multiplicative factor $F$
  \cite{McIntyre1966}. We have measured $F = 7.6$ for our APDs.}\BibitemShut
  {Stop}%
\bibitem [{\citenamefont {Zeiher}\ \emph {et~al.}(2020)\citenamefont {Zeiher},
  \citenamefont {Wolf}, \citenamefont {Isaacs}, \citenamefont {Kohler},\ and\
  \citenamefont {Stamper-Kurn}}]{Zeiher2020}%
  \BibitemOpen
  \bibfield  {author} {\bibinfo {author} {\bibfnamefont {J.}~\bibnamefont
  {Zeiher}}, \bibinfo {author} {\bibfnamefont {J.}~\bibnamefont {Wolf}},
  \bibinfo {author} {\bibfnamefont {J.~A.}\ \bibnamefont {Isaacs}}, \bibinfo
  {author} {\bibfnamefont {J.}~\bibnamefont {Kohler}},\ and\ \bibinfo {author}
  {\bibfnamefont {D.~M.}\ \bibnamefont {Stamper-Kurn}},\ }\href@noop {}
  {\bibinfo {title} {Tracking evaporative cooling of a mesoscopic atomic
  quantum gas in real time}} (\bibinfo {year} {2020}),\ \Eprint
  {https://arxiv.org/abs/2012.01280} {arXiv:2012.01280 [cond-mat.quant-gas]}
  \BibitemShut {NoStop}%
\bibitem [{\citenamefont {Thomas}(2021)}]{GitHub}%
  \BibitemOpen
  \bibfield  {author} {\bibinfo {author} {\bibfnamefont {R.}~\bibnamefont
  {Thomas}},\ }\href
  {https://github.com/kjaergaard-lab/faraday-number-stabilisation} {\bibinfo
  {title} {Software and hardware code for atom number stabilisation via faraday
  rotation.}},\ \bibinfo {howpublished}
  {https://github.com/kjaergaard-lab/faraday-number-stabilisation} (\bibinfo
  {year} {2021})\BibitemShut {NoStop}%
\bibitem [{\citenamefont {McIntyre}(1966)}]{McIntyre1966}%
  \BibitemOpen
  \bibfield  {author} {\bibinfo {author} {\bibfnamefont {R.}~\bibnamefont
  {McIntyre}},\ }\bibfield  {title} {\bibinfo {title} {Multiplication noise in
  uniform avalanche diodes},\ }\href {https://doi.org/10.1109/T-ED.1966.15651}
  {\bibfield  {journal} {\bibinfo  {journal} {IEEE Transactions on Electron
  Devices}\ }\textbf {\bibinfo {volume} {ED-13}},\ \bibinfo {pages} {164}
  (\bibinfo {year} {1966})},\ \bibinfo {note} {conference Name: IEEE
  Transactions on Electron Devices}\BibitemShut {NoStop}%
\end{thebibliography}
\end{document}